\documentclass[twocolumn,twocolappendix,iop,numberedappendix,appendixfloats]{openjournal}

\usepackage[utf8]{inputenc}

\usepackage[british]{babel}

\usepackage{amsmath}
\usepackage{amssymb}
\usepackage{amsfonts}
\usepackage{graphicx}
\usepackage{bm}
\usepackage{natbib}
\usepackage[colorlinks,allcolors=blue]{hyperref}

\usepackage{subfigure}
\usepackage{xspace}
\usepackage{multirow}
\usepackage{caption}
\usepackage{subcaption}
\usepackage{fontawesome}
\usepackage{placeins}

\newcommand{\IAEmu}{\textsc{IAEmu}\xspace}
\newcommand{\halotoolsia}{\textsc{halotools-IA}\xspace}
\newcommand{\mucen}{\mu_\text{cen}\xspace}
\newcommand{\musat}{\mu_\text{sat}\xspace}
\newcommand{\diffhodia}{\textsc{diffHOD-IA}\xspace}
\newcommand{\ncen}{N_\text{cen}\xspace}
\newcommand{\nsat}{N_\text{sat}\xspace}
\newcommand{\githubmaster}{\href{https://github.com/snehjp2/diffHOD-IA}{\faGithub}\xspace}

\begin{document}

\journalinfo{The Open Journal of Astrophysics}
% \submitted{Accepted June 2, 2024}

\shorttitle{Differentiable HOD with Galaxy Intrinsic Alignments}
\shortauthors{S. Pandya et al.}

\title{Differentiable Stochastic Halo Occupation Distribution with Galaxy Intrinsic Alignments}

\author{Sneh Pandya$^{1,2}$}
\author{Jonathan Blazek$^{1}$}

\affiliation{$^{1}$Department of Physics, Northeastern University, Boston, MA 02115, USA}
\affiliation{$^{2}$NSF AI Institute for Artificial Intelligence and Fundamental Interactions (IAIFI)}
\thanks{$^\star$ E-mail: \nolinkurl{pandya.sne@northeastern.edu}}

\begin{abstract}
We present \diffhodia, a fully differentiable implementation of a halo occupation distribution (HOD) model that incorporates galaxy intrinsic alignments (IA). 
Motivated by the \textsc{diffHOD} framework of \citet{horowitz2022}, we create a new implementation that extends differentiable galaxy population modeling to include the orientation-dependent statistics crucial for weak gravitational lensing analyses.
Our implementation combines this HOD formulation with the intrinsic alignment model of \citet{vanalfen_2023}, enabling end-to-end automatic differentiation from HOD and IA parameters through to the galaxy field.
We additionally extend this framework to differentiably model two-point correlation functions, including galaxy clustering and IA statistics.
We validate \diffhodia against the reference \textsc{halotools-IA} implementation using the Bolshoi-Planck simulation, demonstrating agreement within sample variance in galaxy number counts and at $< 2\%$ error in two-point statistics.
We verify the accuracy of gradients computed via automatic differentiation by comparing against finite difference estimates for both HOD and IA parameters. 
We present science use cases that leverage gradients in the parameter space to find the IA of a galaxy field representative of the \textsc{tng300} simulation.
We further use \diffhodia in a Hamiltonian Monte Carlo analysis and compare performance with \halotoolsia and a neural-network based emulator of \halotoolsia, named \IAEmu.
Unlike emulator-based approaches for statistics, \diffhodia provides differentiability at the catalog level, enabling its integration into field-level inference pipelines and extension to arbitrary summary statistics for next-generation weak lensing analyses.  \githubmaster
\end{abstract}


\maketitle

\section{Introduction}
The spatial distribution and intrinsic shapes of galaxies encode fundamental information about the formation and evolution of cosmic structure.
Galaxy clustering, typically quantified through correlation functions, has long served as a primary probe of cosmological parameters and the galaxy-halo connection \citep{Peebles1980, Davis1983}.
More recently, the intrinsic alignments (IA) of galaxy shapes have emerged as both a significant systematic for weak gravitational lensing measurements and a cosmological signal in their own right \citep{Troxel2015, Joachimi2015, Kiessling2015, Blazek_2019}.
As Stage IV surveys such as the Vera C. Rubin Observatory Legacy Survey of Space and Time \citep{lsst}, \textit{Euclid} \citep{euclid}, and the Nancy Grace Roman Space Telescope \citep{roman} come online, precise modeling of both galaxy clustering and IA is essential for extracting unbiased cosmological constraints from weak lensing data.

IA modeling has traditionally relied on analytic approaches rooted in perturbation theory, including the nonlinear alignment model \citep{Hirata_2004,Bridle_2007}, the tidal alignment and tidal torquing (TATT) model \citep{Blazek_2015, Blazek_2019}, and more recent effective field theory treatments \citep{vlah_2020, vlah_2021}.
While these analytic models provide valuable physical insight and computational efficiency, they often struggle to accurately capture nonlinear effects on small scales.
Simulation-based approaches offer a complementary path forward.
Fully nonlinear scales can be described with a halo model \citep{fortuna21a}, which provides a framework for connecting galaxies to their host dark matter halos without explicit modeling of baryonic physics.
This is in addition to semi-analytic models, which combine perturbation theory approaches with halo-based methods \citep{Maion_2024}.
Hydrodynamic simulations provide direct predictions for galaxy alignments, although measurements exhibit significant variance across simulation suites depending on the sub-grid physics employed \citep{tenneti16, samuroff21, vanheukelum2026intrinsicalignmentdisksellipticals}.
For this reason, simulation frameworks that capture the galaxy-halo connection without strong dependence on subgrid physics can be valuable.

The halo occupation distribution (HOD) framework provides a powerful statistical description of the galaxy-halo connection \citep{Peacock2000, Seljak2000, Berlind2002, Zheng2005, Zheng_2007}.
By specifying the probability that a halo of given mass hosts a certain number of galaxies, HOD models can populate dark matter halos from
$N$-body simulations with realistic galaxy distributions.
This approach has been widely successful in modeling galaxy clustering across a range of scales and redshifts \citep{Zehavi2011, Coupon2012, Parejko2013}.
Extensions to the basic HOD framework have incorporated galaxy properties such as color, luminosity, and stellar mass \citep{Zheng2009, Zehavi2011}, as well as secondary halo properties beyond mass \citep{Hearin2016, Wechsler2018}.
Importantly, a growing body of observational evidence from photometric, spectroscopic, and narrowband surveys has established that IA is strongly correlated with these same galaxy properties \citep{Johnston2019, Joachimi2013a, fortuna21a, Fortuna2025, Samuroff2023, Georgiou2025, Siegel2025, NavarroGirones2025}, suggesting that these HOD frameworks are naturally well-suited to model these dependencies compared to analytic approaches such as NLA and TATT.

\citet{Schneider2010} first introduced a halo model for IA, establishing the foundation for connecting galaxy alignments to their host dark matter halos.
Building on this framework, \citet{Joachimi2013a, Joachimi2013b} developed simulation-based approaches that assign galaxy orientations by sampling from distributions encoding alignment with halo shapes, while \citet{Hoffmann_2022} extended this methodology to incorporate alignment with the local tidal field.
The \textsc{halotools} package \citep{Hearin2017} provides a flexible framework for HOD modeling, which was extended by \citet{vanalfen_2023} to include IA modeling via the Dimroth--Watson distribution.
This \halotoolsia framework parameterizes the alignment strength of central and satellite galaxies separately, enabling joint modeling of galaxy clustering and orientation statistics.

A key limitation of the HOD is its reliance on stochastic sampling procedures that are not differentiable, restricting Bayesian inference with HOD simulations to methods like Markov Chain Monte Carlo (MCMC).
This precludes the use of gradient-based optimization and inference methods like Hamiltonian Monte Carlo (HMC; \citealt{Duane1987, Neal2011}), which have proven highly efficient in machine learning and increasingly in cosmological applications (see \citet{dvorkin2022machine} for a review).
As Stage IV survey analyses begin to require inference over larger parameter spaces to account for systematics, MCMC-based inference becomes increasingly computationally demanding in high-dimensional settings.
For instance, the Dark Energy Survey Year 6 result included inference over 50 nuisance parameters, in addition to cosmological parameters \citep{descollaboration2026darkenergysurveyyear}.
In \citet{Piras_2024}, it was estimated that a Stage IV full 3x2 analysis would require up to 12 years of compute time on 48 CPU cores with MCMC-like sampling, which is contrasted with 8 days on GPU with differentiable sampling techniques.
Consequently, there is a growing need for more computationally efficient (differentiable) methods that enable tractable inference over many free parameters.

One path forward is training NN-based emulators, which have exhibited significant forward modeling and inference speed-ups in modeling IA in hydrodynamic simulations \citep{Jagvaral_2022, jagvaral2023diffusion, jagvaral24_arxiv} and in the HOD framework \citep{Pandya_2025}.
Yet another path is through differentiable simulations, which enable the computation of gradients of observables with respect to model parameters via automatic differentiation.
This enables gradient descent optimization, HMC sampling, and integration into differentiable analysis pipelines.

\citet{horowitz2022} introduced \textsc{diffHOD}, a differentiable implementation of the HOD framework that employs continuous relaxations of discrete sampling distributions.
By utilizing the Gumbel-Softmax trick \citep{jang2017categoricalreparameterizationgumbelsoftmax, Maddison2017} for Bernoulli and Poisson distributions, \textsc{diffHOD} enables end-to-end gradient flow from HOD parameters through to galaxy catalogs.
This approach was shown to accelerate parameter inference by orders of magnitude compared to traditional likelihood-free methods.
Separately, \citet{Hearin_2022} developed differentiable estimators for galaxy clustering statistics, enabling gradients to flow through two-point correlation function measurements.

In this work, we present \diffhodia, which extends the differentiable HOD framework to include galaxy intrinsic alignments.
Our implementation combines the \citet{Zheng_2007} HOD formulation with the intrinsic alignment model of \citet{vanalfen_2023}, enabling automatic differentiation from HOD and IA parameters through to the galaxy field and summary statistics.
We develop a differentiable sampling procedure for the Dimroth--Watson distribution via inverse cumulative distribution function methods, and implement differentiable estimators for the galaxy position-position, position-orientation and orientation-orientation correlation functions.
We validate \diffhodia against the reference \halotoolsia implementation using the Bolshoi-Planck simulation \citep{Klypin2016, Rodriguez-Puebla2016}, demonstrating excellent agreement in galaxy number counts and two-point statistics.
We verify gradient accuracy by comparison with finite difference estimates.
We then demonstrate science applications including gradient-based optimization to recover IA parameters from mock observations and HMC inference that achieves substantial speedups over traditional MCMC approaches.

The structure of this paper is as follows.
In Section \ref{sec:originalformulation}, we review the \halotoolsia formulation for HOD modeling with intrinsic alignments.
Section \ref{sec:diffhodia} describes our differentiable implementation, including the relaxed sampling procedures and differentiable Dimroth--Watson sampling.
Section \ref{sec:diff_correlations} presents our differentiable correlation function estimators for IA statistics.
In Section \ref{sec:accuracy}, we validate the accuracy of \diffhodia and its gradients.
Section \ref{sec:experiments} demonstrates gradient-based optimization and HMC inference applications.
We summarize and discuss future directions in Section \ref{sec:summary}.

\section{The \texttt{halotools-IA} algorithm}
\label{sec:originalformulation}
We begin by reviewing the original \halotoolsia formalism \citep{vanalfen_2023, VanAlfen2025}, which builds upon the halo occupation distribution (HOD) framework \citep{Peacock2000, Seljak2000, Berlind2002}, adopting the commonly used parameterization introduced by \citet{Zheng2005}.
The \halotoolsia model extends this to seven free parameters, with five of them governing the halo occupation and two of them governing the galaxy IA. These parameters are as follows:
\begin{equation}
\{
\underbrace{\log{M_{\text{min}}}, \sigma_{\log{M}}, \log{M_{0}}, \log{M_{1}}, \alpha}_{\text{HOD}},
\underbrace{\mu_\text{cen},\mu_\text{sat}}_{\text{IA}}
\}.
\end{equation}
The first two HOD parameters, $\log{M_{\text{min}}}$ and $\sigma_{\log{M}}$, are related to the occupation of central galaxies, while $\log{M_{0}}$, $\log{M_{1}}$, and $\alpha$ govern satellite galaxy occupation.
More detailed descriptions of the HOD parameters can be found in \citet{Zheng_2007}. 
The IA parameterization is a two-parameter family that statistically describes the alignment strength of central and satellite galaxies with respect to their host halos, as defined in \citet{vanalfen_2023}.

\subsection{Central Occupation}
\label{sec:central_occ}

The central galaxy occupation is characterized by the minimum halo mass required to host a central galaxy, and the width of the transition around this threshold.
These are described by the parameters $\log{M_{\text{min}}}$ and $\sigma_{\log{M}}$.
The expected number of central galaxies in a halo of mass $M$ is given by:
\begin{equation}
\label{eqn:ncen}
    \langle N_{\mathrm{cen}}(M) \rangle =
\frac{1}{2}\left[ 1 + \operatorname{erf}\!\left( 
\frac{\log M - \log M_{\min}}{\sigma_{\log M}} 
\right) \right],
\end{equation}
where $\operatorname{erf}$ denotes the error function. 
Since the presence of a central galaxy is a binary outcome, central galaxies are assigned to host halos via a Bernoulli distribution:
$$
N_\text{cen} \sim \text{Bernoulli}(p = \langle N_{\mathrm{cen}}(M) \rangle).
$$

\subsection{Satellite Occupation}
\label{sec:satellite_occ}

Satellite galaxy occupation follows a power-law model, governed by the parameters $\log{M_{0}}$, $\log{M_{1}}$, and $\alpha$. 
The expected number of satellite galaxies in a halo of mass $M$ is defined as:
\begin{equation}
\label{eqn:nsat}
\langle N_{\mathrm{sat}}(M) \rangle
= \langle N_{\mathrm{cen}}(M) \rangle
\left( \frac{M - M_{0}}{M_{1}} \right)^{\alpha},
\end{equation}
indicating that satellites are only present in halos that already host a central galaxy. 
Because multiple satellite galaxies can inhabit a single halo, they are sampled via a Poisson distribution:
\begin{equation}
N_\text{sat} \sim \text{Poisson}(\lambda = \langle N_{\mathrm{sat}}(M) \rangle).
\end{equation}
We utilize a \texttt{SubhaloPhaseSpace} model for the IA in which satellite galaxies are always placed at the center of subhalos.
These are preferentially placed in more massive subhalos first.
In the event that the number of satellites exceeds the number of available subhalos, the satellite occupation resorts to a \texttt{NFWPhaseSpace} mode, where the remaining satellites are spatially distributed by sampling from a Navarro-Frenk-White (NFW) distribution \citep{Navarro_1996} about the central halo.
One can also utilize the \texttt{NFWPhaseSpace} for all satellite occupation, in the event that subhalo information is not available.
\label{sec:HOD}

\subsection{Galaxy Intrinsic Alignments}
\label{sec:galaxy_IA}

Both galaxies and (sub)halos are modeled as triaxial homologous ellipsoids, meaning the orientations of the halos and galaxies can be described entirely from their axes.
The misalignment angle of the galaxies with their host halos, $\theta_{\text{MA}}$, is governed by sampling from a Dimroth--Watson distribution \citep{Watson:1965:EDS}.
The Dimroth--Watson distribution is chosen to model galaxy alignments as it provides a maximum entropy distribution over a sphere, while accounting for the spin-2 symmetry of galaxy orientations.
The probability distribution function (PDF) for the Dimroth--Watson $P(\theta, \phi)$ is defined as
\begin{equation}
\label{eqn:dimrothwatson}
P(\theta, \phi) = \frac{B(\kappa)}{2\pi} \, e^{-\kappa \cos^{2}(\theta)} \sin(\theta)\, d\theta\, d\phi,
\end{equation}
for polar angle $\theta = \theta_{\text{MA}}$ and azimuthal angle $\phi$. The normalization factor is given by
\begin{equation}
\label{eqn:normalization}
B(\kappa) = \frac{1}{2} \int_{0}^{1} e^{-\kappa t^{2}} \, dt,
\end{equation}
where $t=\cos(\theta)$.
In this formulation, $\phi$ is sampled from a uniform distribution.

The fundamental parameter governing galaxy and (sub)halo alignment is $\kappa$. It is convenient to reparameterize this as
\begin{equation}
    \mu = \frac{-2 \tan^{-1}(\kappa)}{\pi},
\end{equation}
such that $\mu = \pm1$ corresponds to perfect (mis)alignment, and $\mu = 0$ corresponds to random alignments.
Both central and satellite galaxies are assigned their own alignment strengths, $\mucen$ and $\musat$, respectively.
\halotoolsia includes two separate alignment formulations for satellites: \emph{subhalo alignment}, where satellites are oriented with respect to their host subhalo, and \emph{radial alignment}, where satellites are oriented with respect to the host halo radial vector.
The radial alignment model also admits two alignment strength possibilities: constant alignment strength, where it is the same for all galaxies, and distance-dependent alignment strength, where it has a power-law dependence on the distance to the central galaxy.
More generally, the alignment strength for each galaxy can be determined based on galaxy properties.
We implement all aforementioned cases in the \diffhodia code; in this paper, all experiments utilize the radial alignment model with constant alignment strength.

\section{Differentiable \texttt{halotools-IA}}
\label{sec:diffhodia}

\begin{figure*}
    \centering
    \includegraphics[width=\textwidth]{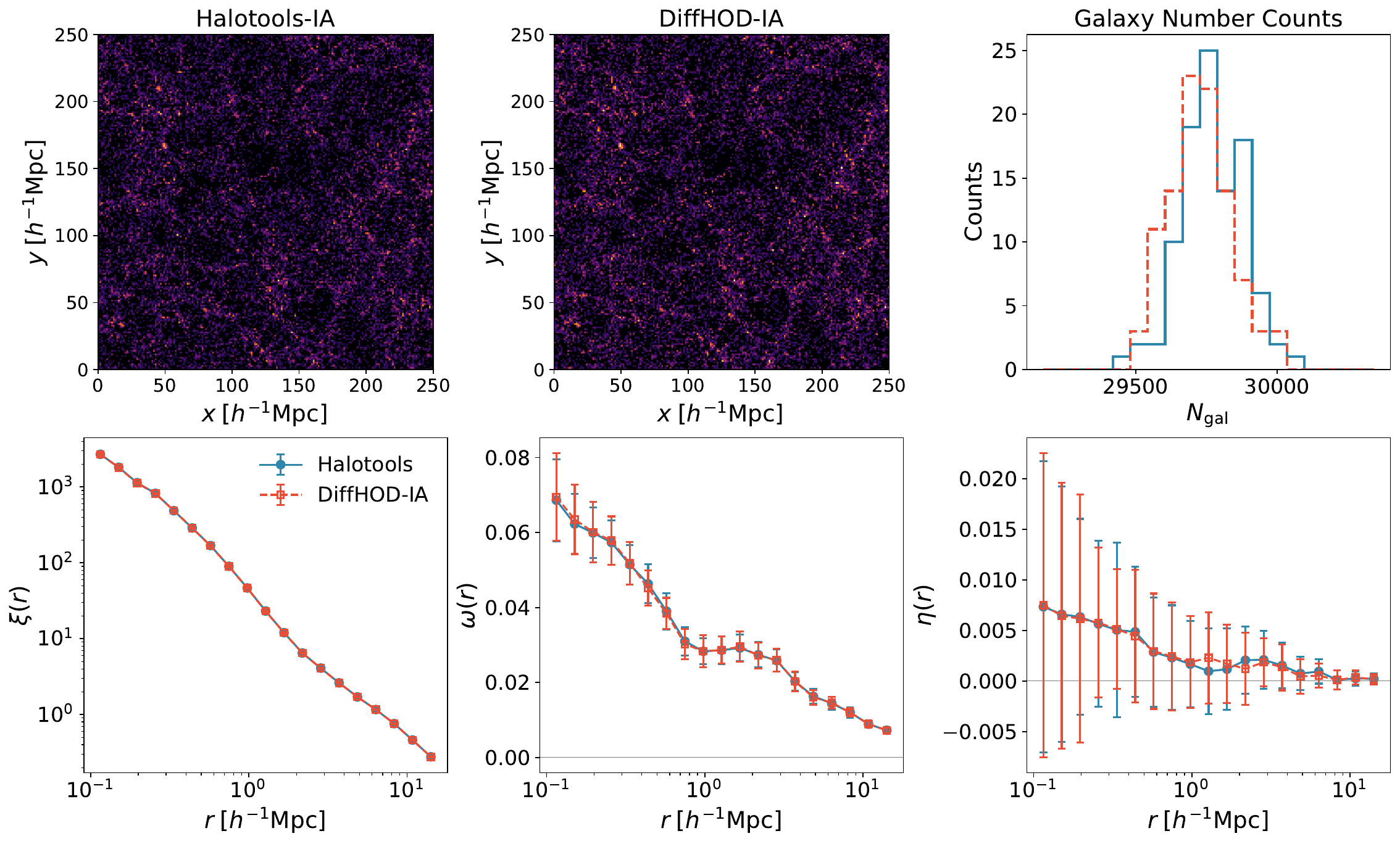}
    \caption{Validation of \diffhodia against the reference \halotoolsia implementation for the \textsc{tng300} fiducial HOD. \textbf{Top left and center:} Projected galaxy density fields across the simulation volume along the line of sight. Both implementations produce visually indistinguishable large-scale structure. \textbf{Top right:} Distribution of galaxy number counts $N_\text{gal}$ across 100 realizations using identical random seeds. Both implementations produce consistent galaxy number densities with similar scatter. \textbf{Bottom left:} Galaxy position-position correlation function $\xi(r)$ averaged over 100 realizations, with error bars indicating the standard deviation across realizations. The two implementations show excellent agreement across all scales. \textbf{Bottom center and right:} Galaxy position-orientation correlation function $\omega(r)$ and orientation-orientation correlation function $\eta(r)$. These correlations show strong agreement between the two implementations across all scales. $\eta(r)$ exhibits larger statistical noise and error bars due to the effects of galaxy shape noise. Despite the noise, the two implementations remain consistent within uncertainties.}
    \label{fig:fullcomparison}
\end{figure*}

We now proceed to outline the differentiable HOD methodology that includes galaxy IA, henceforth referred to as \diffhodia. In particular, \diffhodia allows an end-to-end differentiable mapping for:
\begin{align}
        \text{HOD} \; + \text{IA parameters} &\rightarrow \text{misaligned galaxy field} \nonumber \\&\rightarrow \text{summary statistics} \nonumber
\end{align}
under the HOD model of \citet{Zheng_2007} and IA model of \citet{vanalfen_2023}.
Our contributions also include differentiable modeling of two-point correlation functions (2PCFs) within this framework, following the work of \cite{Hearin_2022}.
Differentiable computation of summary statistics is \textsl{a priori} nontrivial, because common cosmological summaries such as the 2PCF rely on inherently discrete operations like galaxy pair counting.
We will showcase the utility of \diffhodia with both 2PCFs and other objectives.

For galaxy clustering, our differentiable HOD methodology closely follows that of \textsc{diffHOD} \citep{horowitz2022}, which we will summarize below.
\textsc{diffHOD} was shown to agree with (non-differentiable) HOD-generated catalogs at the level of the power spectrum to within $1\%$ at the scales considered ($k < 3 \; h/\text{Mpc}$), which is generally sufficient for Stage IV cosmological analyses.
We extend this framework by introducing a differentiable procedure for Dimroth--Watson sampling, employing inverse cumulative distribution function (CDF) sampling.
As we adopt their methodology closely, we refrain from extensively benchmarking 
the HOD component of \diffhodia, and instead focus our analyses on the IA modeling and its accuracy compared to \halotoolsia.

\subsection{Differentiable Sampling}

Sampling from distributions is not an operation whose gradients can be tracked, as an individual sample $z$ does not encode parametric information about the distribution which it was sampled from. 
A common approach for backpropagating through distributions is via the \emph{reparameterization} trick, as is extensively used in variational autoencoders \citep{kingma2022autoencodingvariationalbayes}.
In this procedure, the random variable is expressed as a deterministic function of both the distribution parameters and a source of parameter-free noise. 
For example, for a normally distributed variable $z \sim \mathcal{N}(\mu, \sigma^2)$, one instead samples $\epsilon \sim \mathcal{N}(0, 1)$ and rewrites $z$ as:
\begin{equation}
z = \mu + \sigma \cdot \epsilon,
\end{equation}
which allows gradients to be backpropagated through $\mu$ and $\sigma$.

Discrete distributions, such as the Bernoulli or Poisson distribution, assign binary values (e.g., galaxy/no galaxy) to their random variables. 
As a result, it is difficult to differentiably sample from such distributions. 
One approach to this challenge is the Gumbel-Softmax trick \citep{jang2017categoricalreparameterizationgumbelsoftmax}, which defines a continuous relaxation of a discrete distribution, allowing gradients to be backpropagated via the reparameterization trick.

Specifically, for a categorical distribution with class probabilities $\{\pi_1, \dots, \pi_k\}$, a sample $z$ is typically drawn as a one-hot vector using:
\begin{equation}
    z \sim \mathrm{Categorical}(\pi_1, \dots, \pi_k),
\end{equation}
where $z_i = 1$ indicates the selected category. 
However, this process is non-differentiable due to the discrete $\mathrm{argmax}$ operation implicit in categorical sampling and the inherently stochastic nature of sampling.
To enable differentiability, one introduces the reparameterization trick by injecting independent Gumbel noise $g_i \sim \mathrm{Gumbel}(0,1)$---a choice that follows from the Gumbel-Max trick for sampling from categorical distributions \citep{Maddison2017}---and replacing the $\mathrm{argmax}$ with a differentiable $\mathrm{softmax}$ approximation. 
The relaxed sample $y_i$ is then given by:
\begin{equation}
y_i = \frac{\exp\left((\log \pi_i + g_i)/\tau\right)}{\sum_{j=1}^k \exp\left((\log \pi_j + g_j)/\tau\right)},
\end{equation}
where $\tau > 0$ is a temperature parameter that controls the degree of approximation. 
In the limit $\tau \to 0$, the softmax converges to a hard (non-differentiable) categorical sample; for larger $\tau$, the distribution becomes smoother and more uniform.
The trade off is in the gradients, where small $\tau$ values result in a large variance of gradients, while large $\tau$ results in a smaller variance.

\subsection{Differentiable Central Occupation}

Central occupation sampling is defined by a Bernoulli distribution, as described in Section \ref{sec:central_occ}.
We must differentiably sample
\begin{equation}
N_\text{cen} \sim \text{Bernoulli}(p = \langle N_{\mathrm{cen}}(M \;| \; M_{\text{min}}, \sigma_{\log{M}}) \rangle).
\end{equation}
To this end, we utilize the Gumbel-Softmax trick in defining the Relaxed Bernoulli distribution:
\begin{equation}
\label{eqn:relaxed_bernoulli}
N_{\mathrm{cen}} = \frac{1}{1 + \exp\left(-\left( \log\left( \frac{p}{1 - p} \right) + \epsilon \right)/\tau \right)}
\end{equation}
with $\epsilon \sim \mathrm{Logistic}(0, 1)$.
We adopt a temperature value of $\tau = 0.1$, consistent with the full analysis of the occupation accuracy dependent on values of $\tau$ in \citet{horowitz2022}. 

\subsection{Differentiable Satellite Occupation}

Satellite occupation requires sampling from a Poisson distribution as
\begin{equation}
N_\text{sat} \sim \text{Poisson}(\lambda = \langle N_{\mathrm{sat}}(M \;| \; M_0, M_1, \alpha) \rangle).
\end{equation}
In \citet{horowitz2022}, it was proposed to treat each potential satellite galaxy assignment as an independently sampled Bernoulli distribution with probability $p = \lambda/N_{\text{max}}$, where $N_{\text{max}}$ is the number of trials and $\lambda$ is the Poisson rate.
The resulting statistics are then Binomial distributed, which has an identical mean to a Poisson distribution.
The two distributions only differ in the variance, where
\begin{align}
\mathrm{Var}(N_{\mathrm{sat}}^{\mathrm{Pois.}}) 
&= \langle N_{\mathrm{sat}} \rangle \\
\mathrm{Var}(N_{\mathrm{sat}}^{\mathrm{Bin.}}) 
&= \langle N_{\mathrm{sat}} \rangle \left( 1 - \frac{\langle N_{\mathrm{sat}} \rangle}{N_{\text{max}}} \right) \label{eq:binomial_variance}.
\end{align}

In the limit $N_{\text{max}} \to \infty$, the two distributions are identical in their first two moments.
In this work, we use a fiducial value of $N_{\text{max}} = 48$ for all experiments, in line with the results of \citet{horowitz2022}.

We have now simplified satellite sampling into independent Bernoulli sampling identical to Equation \ref{eqn:relaxed_bernoulli} with $p = \langle N_{\mathrm{sat}} \rangle/N_{\text{max}}$.  
In this work, all experiments will utilize the \texttt{SubhaloPhaseSpace}, wherein satellite galaxies are deterministically placed at the centers of subhalos, prioritized by subhalo mass (specifically, \texttt{halo\_mpeak}).

This prioritization is implemented differentiably by applying a softmax over the subhalos within each host halo, with logits determined by subhalo \emph{rank}, i.e., the ordering of subhalos by decreasing mass within each host halo.
Formally, the rank-weighted soft assignment $q_i$ for subhalo $i$ is computed as:
\begin{equation}
\label{eqn:softmaxweight}
q_i = \frac{\exp\left(-\mathrm{rank}_i / t_{\mathrm{rank}}\right)}{\sum_{j \in \mathcal{H}_h} \exp\left(-\mathrm{rank}_j / t_{\mathrm{rank}}\right)},
\end{equation}
where $\mathrm{rank}_i \in \{0, 1, 2, \dots\}$ is the indexed position of subhalo $i$ within host halo $h$, $\mathcal{H}_h$ is the set of all subhalos associated with that host, and $t_{\mathrm{rank}} = 0.5$ is a temperature parameter controlling the sharpness of the prioritization.
These normalized weights $q_i$ are then scaled by the expected number of satellites $\langle N_{\mathrm{sat}} \rangle$ to yield the relaxed per-subhalo satellite probabilities.

\subsection{Differentiable NFW Sampling}
\label{sec:differentiable_nfw}

% \jab{No subhalo borrowing? Contrast with Van Alfen.}
In the event that $N_{\text{sat}} > |\mathcal{H}_h|$, we assign the locations of all remaining satellite galaxies according to samples from an NFW distribution. 
We generate satellite positions via an inverse CDF sampling procedure, which we will also use for IA sampling in Section \ref{sec:diffIA}. 
Our treatment follows the approach of \citet{horowitz2022}, but rather than using their closed-form approximation based on the Lambert $W$ function, we employ a numerically stable Newton iteration to invert the CDF.
Both approaches are differentiable and similarly accurate.
We begin with the CDF of the NFW profile:
\begin{equation}
P(<r) = \frac{\ln(1 + c r/r_{\mathrm{vir}}) - c r/r_{\mathrm{vir}} / (1 + c r/r_{\mathrm{vir}})}{\ln(1 + c) - c / (1 + c)},
\end{equation}
where $c$ is the concentration parameter of the host halo and $r_{\mathrm{vir}}$ is the virial radius of the host halo.
We obtain position samples by drawing $u \sim \text{Uniform}(0,1)$ and solving for $r = P^{-1}(u)$ via Newton's method for six iterations, which was found to be sufficient yielding a convergence error of $\lesssim 10^{-4}$.

\subsection{Differentiable Galaxy Intrinsic Alignments}
\label{sec:diffIA}

The central and satellite galaxy misalignments are modeled by the Dimroth–Watson distribution, defined by its PDF in Equation~\ref{eqn:dimrothwatson}. 
To differentiably sample from it, we perform inverse CDF sampling, analogous to the NFW sampling procedure described in Section~\ref{sec:differentiable_nfw}.

The original (unnormalized) marginal density over the misalignment polar angle \( \theta \in [0, \pi] \) is given by:
\begin{equation}
    p(\theta) \propto \exp(-\kappa \cos^2 \theta)\, \sin\theta.
\end{equation}
Changing variables to \( t = \cos\theta \in [-1,1] \), the density becomes:
\begin{equation}
    p(t) = \frac{1}{Z(\kappa)} \exp(-\kappa t^2), \qquad t \in [-1, 1],
\end{equation}
with normalization constant
\begin{equation}
Z(\kappa) = \int_{-1}^{1} \exp(-\kappa t^2)\, dt.
\end{equation}
To draw differentiable samples, we define the CDF
\begin{equation}
F(t \mid \kappa) = \frac{1}{Z(\kappa)} \int_{-1}^{t} \exp(-\kappa s^2)\, ds.
\end{equation}
This integral admits closed-form expressions that depend on the sign of $\kappa$, yielding the following piecewise form of the CDF:
\begin{equation}
\label{eqn:allcdfcases}
F(t \mid \kappa) =
\begin{cases}
\frac{1}{2} \left[ 1 + \dfrac{\operatorname{erf}(\sqrt{\kappa} t)}{\operatorname{erf}(\sqrt{\kappa})} \right], & \kappa > 0 \\
\frac{t + 1}{2}, & \kappa = 0 \\
\frac{1}{2} \left[ 1 + \dfrac{\operatorname{erfi}(\sqrt{-\kappa} t)}{\operatorname{erfi}(\sqrt{-\kappa})} \right], & \kappa < 0.
\end{cases}
\end{equation}

To generate samples, we invert the CDF by solving \( t = F^{-1}(u \mid \kappa) \) for \( u \sim \mathrm{Uniform}(0,1) \).
For convenience, we rescale the uniform random variable via the transformation \( u' = 2u - 1 \), which maps \( u \in (0,1) \) to \( u' \in (-1,1) \); we drop the prime in what follows.
Inverting the appropriate branch of Equation~\ref{eqn:allcdfcases} then yields:
\begin{equation}
\label{eqn:inversecdf}
\cos(\theta) =
\begin{cases}
\frac{1}{\sqrt{\kappa}}\, \operatorname{erf}^{-1}\left[ \operatorname{erf}(\sqrt{\kappa}) \cdot u \right], & \kappa > 0 \\
u, & \kappa = 0 \\
\frac{1}{\sqrt{-\kappa}}\, \operatorname{erfi}^{-1}\left[ \operatorname{erfi}(\sqrt{-\kappa}) \cdot u \right], & \kappa < 0
\end{cases}
\end{equation}
from which we can compute $\theta$.
To construct the full 3D orientation vector \( \boldsymbol{n} \), we draw \( u \sim \mathrm{Uniform}(-1,1) \) to obtain \( t = \cos\theta \) using Equation~\ref{eqn:inversecdf}, and independently sample \( \phi \sim \mathrm{Uniform}(0, 2\pi) \). These are converted to Cartesian coordinates as:
\begin{equation}
    \boldsymbol{n} = 
    \begin{bmatrix}
        \sin \theta \cos \phi \\
        \sin \theta \sin \phi \\
        \cos \theta
    \end{bmatrix}.
\end{equation}
While Equation \ref{eqn:inversecdf} admits analytic inverse forms, the implementation evaluates the inverse CDF analytically when possible, and otherwise solves the inverse relation numerically using a small number of Newton iterations when the inverse involves $\text{erfi}^{-1}$, which is not available in standard libraries.
The resulting galaxy orientation vectors are fully differentiable with respect to the IA parameters.

\section{Differentiable Correlation Functions}
\label{sec:diff_correlations}
% \jab{Move this section later?}

Typical cosmological analyses employ low dimensional summary statistics computed over the full galaxy field.
We proceed by outlining a prescription to differentiably calculate 2PCFs, enabling an end-to-end differentiable pipeline from halo occupation and IA parameters to summary statistics.
We find that computing the IA statistics is directly differentiable with respect to the IA parameters using discrete galaxy catalogs; computing $\xi(r)$ requires generating catalogs where the galaxies have a weight proportional to their occupation probability, as done in \citet{Hearin_2022}.
We note that the IA statistics $\omega(r)$ and $\eta(r)$ also depend on the HOD parameters through the galaxy positions and number counts; however, for the experiments in this work, we operate in a fixed HOD setting to validate the differentiable IA implementation.
We develop and benchmark the accuracy for both the weighted $\xi(r)$ and IA correlation estimators in Appendix \ref{app:correlations}.
All correlation measurements use 20 logarithmically-spaced bins from $r = 0.1 \; h^{-1} \text{Mpc}$ to $r = 16 \; h^{-1} \text{Mpc}$.
These scales capture the transition from the 1-halo to 2-halo regime, while also allowing correlations to still be measured quickly.

\subsection{Intrinsic Alignment Statistics}
In this context, there are two IA statistics of interest: $\omega(r)$ and $\eta(r)$.
Our treatment does not currently include full galaxy shape information, and the correlations as defined below are only sensitive to galaxy orientations.
The galaxy position-orientation correlation function $\omega(r)$ is defined as:
\begin{equation}
    \omega(r) = \langle (\hat{e}(\mathbf{x}) \cdot \hat{r})^2 \rangle - \frac{1}{3}, 
\end{equation}
where $\hat{e}$ is the galaxy orientation unit vector that specifies the intrinsic orientation of each galaxy's major axis, $\mathbf{x}$ is the position vector of a galaxy, and $\hat{r}$ is the unit vector in the direction of galaxy separation in 3D.
The galaxy orientation-orientation correlation function $\eta(r)$ is defined as:
\begin{equation}
    \eta(r) = \langle (\hat{e}(\mathbf{x}) \cdot \hat{e}(\mathbf{x} + \mathbf{r}))^2 \rangle - \frac{1}{3}, 
\end{equation}
which measures the correlation between the orientations of galaxy pairs as a function of their separation.
The factor of $1/3$ in these equations account for the fact that
\begin{equation}
    \frac{1}{4\pi}\int_0^{2\pi} \int_0^{\pi} \cos^2\theta \sin \theta \; \text{d}\theta \; \text{d}\phi = \frac{1}{3},
\end{equation}
where integrating $\cos^2(\theta)$ over the sphere corresponds to the case of random alignments.

We pre-compute all neighbor pairs $(i, j)$ with separations $|\mathbf{r}_{ij}| < r_{\rm max}$ using a KD-tree with periodic boundary conditions.
This is implemented via \texttt{scipy}.
Precomputation strategies to avoid redundant pair-finding operations have been employed 
previously to accelerate HOD parameter exploration, including tabulating pair counts as 
a function of halo mass \citep{Neistein_2011, Zheng_2016}.
Our approach differs in that we store the explicit pair indices for a fixed galaxy catalog, 
enabling gradient flow through the orientation-dependent alignment statistics while treating 
the neighbor structure as constant.
For each pair, we compute the separation vector with periodic wrapping. For $\omega(r)$, we compute the position-orientation alignment:
\begin{equation}
\label{eqn:aijomega}
a_{ij}^{\omega} = \left(\hat{e}_i \cdot \hat{r}_{ij}\right)^2.
\end{equation}
For $\eta(r)$, we compute the orientation-orientation alignment:
\begin{equation}
a_{ij}^{\eta} = \left(\hat{e}_i  \cdot \hat{e}_j \right)^2,
\end{equation}
which measures the alignment between the orientation vectors of galaxies $i$ and $j$.

Pairs are assigned to radial bins, and the correlation functions are estimated as:
\begin{equation}
\omega(r_k) = \frac{\sum_{(i,j) \in \mathcal{B}_k} a_{ij}^{\omega}}{|\mathcal{B}_k|} - \frac{1}{3}
\end{equation}
and
\begin{equation}
\eta(r_k) = \frac{\sum_{(i,j) \in \mathcal{B}_k} a_{ij}^{\eta}}{|\mathcal{B}_k|} - \frac{1}{3}.
\end{equation}
Here $\mathcal{B}_k$ denotes the set of all galaxy pairs whose separations fall within radial bin $k$, and $|\mathcal{B}_k|$ is the number of such pairs.
Crucially, since galaxy positions are fixed, the pair counts $|\mathcal{B}_k|$ are constants with respect to the IA parameters.
Gradients flow exclusively through the orientation vectors $\hat{e}_i$, which depend on $\mucen$ and $\musat$ via the differentiable Dimroth--Watson sampling described in Section~\ref{sec:diffIA}.

\section{Accuracy \& Gradients}
\label{sec:accuracy}

\begin{table}
\caption{\textsc{Bolshoi-Planck} simulation parameters.}
\begin{tabular}{ccccccc}
\hline
Particle Mass & $\Omega_{m,0}$ & $\sigma_8$ & $n_s$ & $h$ & $L_{\rm box}$ & $z$ \\
($h^{-1}M_\odot$) &  &  &  &  & ($h^{-1}$Mpc) &  \\
\hline
$\sim 10^8$ & 0.30711 & 0.82 & 0.96 & 0.70 & 250 & 0 \\
\hline
\end{tabular}
\label{tab:bolshoi}
\end{table}

To evaluate \diffhodia, we assess the accuracy of its IA implementation by comparing it against \halotoolsia.
We also examine the model’s differentiability with respect to both HOD and IA parameters, and compare the resulting gradients to those obtained via finite difference methods.
A detailed performance benchmark of the HOD component is not provided, as its implementation closely follows that of \citet{horowitz2022}.
However, we do benchmark the computational cost of \diffhodia\ by comparing $N_{\rm max}$, used in the differentiable satellite occupation, to $N_{\rm Newton}$, used for inverse CDF sampling from the Dimroth--Watson distribution.
These results are presented in Appendix \ref{app:efficiency}.

We will construct a mock observable galaxy catalog and use this fiducial catalog as a reference in the following sections.
This catalog is constructed using the parameters:
\begin{align*}
&\log M_{\min} = 12.54, \;
\sigma_{\log M} = 0.26, \;
\log M_{0} = 12.68 \nonumber \\ \nonumber
&\log M_{1} = 13.48, \;
\alpha = 1.0, \; \mu_{\text{cen}} = 0.79, \; \mu_{\text{sat}} = 0.30.
\end{align*}
The HOD parameters were determined by fitting the halo occupation to match a galaxy catalog from the \textsc{tng300} \citep{nelson2021illustristng} simulation for a stellar mass $M_*$ cutoff of $\log M_* \geq 10.5$.
This HOD is run on the Bolshoi-Planck dark matter catalog at $z=0$, whose simulation parameters can be found in Table \ref{tab:bolshoi}.
We adopt the HOD configuration used in \citet{vanalfen_2023, Pandya_2025}; details of how the catalog was constructed to match \textsc{tng300} are given in \citet{vanalfen_2023}, and the IA parameters correspond to the \textsc{tng300} best-fit values reported in \citet{Pandya_2025}.

\subsection{Comparison to \halotoolsia}
\label{sec:comparisonhalotools}

We compare the performance between \diffhodia and \halotoolsia using visualizations of the galaxy field density, comparison of 1-pt statistics (i.e., galaxy number counts, $N_\text{gal}$), and relevant 2-pt statistics.
This is shown in Figure \ref{fig:fullcomparison}.
This comparison provides a comprehensive metric for the performance of \diffhodia.

We find excellent agreement between \diffhodia and \halotoolsia for the fiducial galaxy catalog used.
The galaxy field densities produced by the two simulations are visually consistent, as shown in the top left and center panels of Figure \ref{fig:fullcomparison}.
As galaxy placements are deterministic due to the use of \texttt{SubhaloPhaseSpace}, this is representative of agreement in the per-halo $\langle N_{\rm cen}\rangle$ and $\langle N_{\rm sat}\rangle$ between the two implementations.
In both cases, approximately $0.17\%$ of the remaining satellites were occupied according to \texttt{NFWPhaseSpace}.

To go beyond a visual comparison, this is further confirmed upon examining the histogram of $N_\text{gal}$ across 100 realizations in the top right panel of Figure \ref{fig:fullcomparison}.
The \diffhodia galaxy number counts are $29772 \pm 109$, and the \halotoolsia number counts are $29729 \pm 110$, illustrating excellent agreement.
Minor differences in the mean values may stem from the fact that the seed mapping between \diffhodia and \halotoolsia is not one-to-one, which can lead to small deviations over a finite number of realizations.

In the bottom panels of Figure \ref{fig:fullcomparison}, we compare $\xi(r)$, $\omega(r)$, and $\eta(r)$ between the two implementations.
This is using the (non-differentiable) \halotoolsia implementation for the correlation estimators, to highlight any potential differences at the 2PCF level coming solely from the galaxy occupation and IA.
We find generally excellent visual agreement between \diffhodia and \halotoolsia across all three statistics.
This even includes the IA statistic $\eta(r)$, which is much noisier than $\omega(r)$ due to galaxy shape noise.
We additionally see comparable error bars between the implementations across 100 realizations.
This is confirmed upon inspecting the fractional error of the correlations, from which we find a mean bias of $0.28\%$ for $\xi(r)$, $0.16\%$ for $\omega(r)$, and $1.77\%$ for $\eta(r)$.
We benchmark the accuracy of the differentiable correlation estimators as outlined in Section \ref{sec:diff_correlations} in Appendix \ref{app:correlations}.

\subsection{HOD Gradients in \diffhodia}

\begin{figure*}
    \centering
    \includegraphics[width=0.9\textwidth]{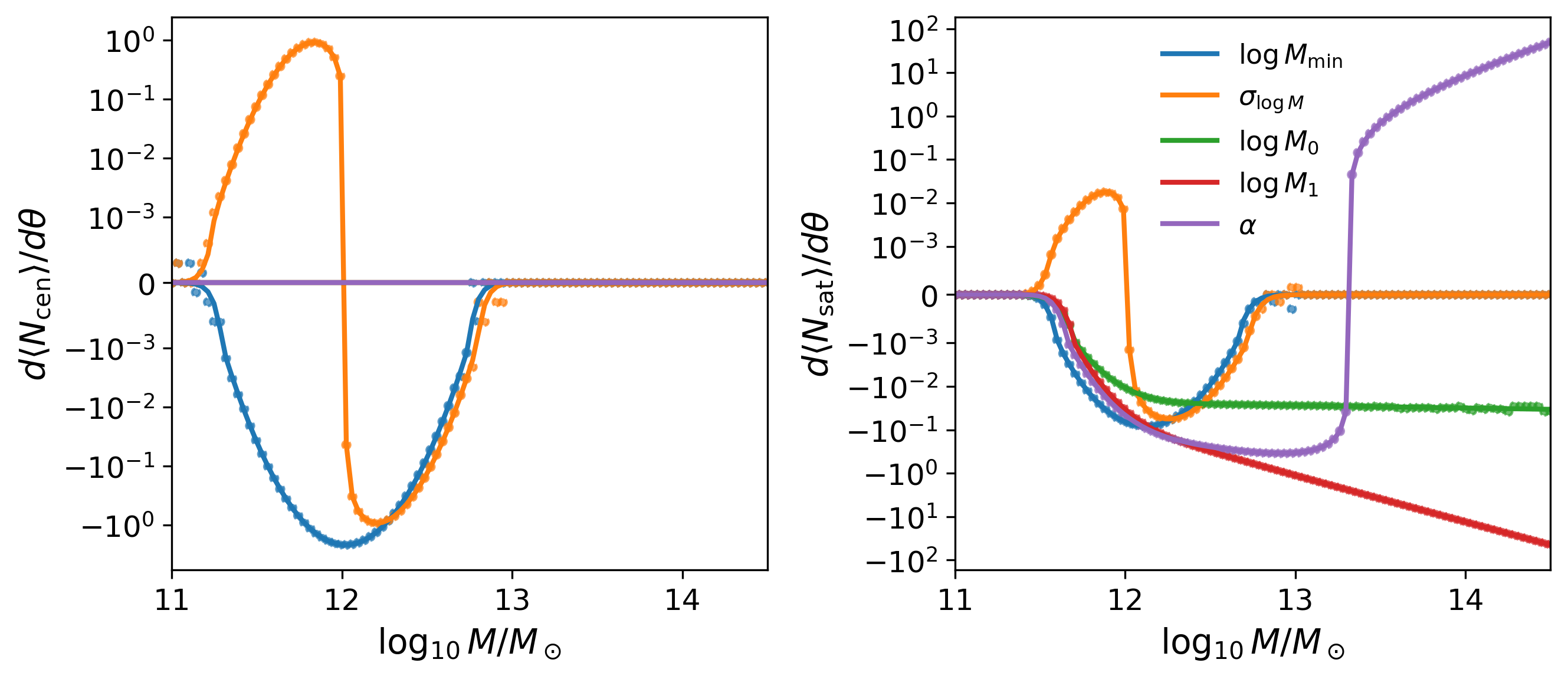}
    \caption{Gradients of the halo occupation distribution functions with respect to HOD parameters as a function of halo mass. \textbf{Left panel:} Gradients of the mean central galaxy occupation $\langle N_{\rm cen} \rangle$ with respect to $\log M_{\rm min}$ (blue) and $\sigma_{\log M}$ (orange). The gradients are largest in the transition region around $\log_{10} M/M_\odot \approx 12$ where the occupation probability transitions from 0 to 1, and vanish at high masses where $\langle N_{\rm cen} \rangle$ saturates to unity. \textbf{Right panel:} Gradients of the mean satellite galaxy occupation $\langle N_{\rm sat} \rangle$ with respect to all five HOD parameters: $\log M_{\rm min}$, $\sigma_{\log M}$, $\log M_0$, $\log M_1$, and $\alpha$. In both panels, solid lines show gradients computed via automatic differentiation and dotted points show finite difference estimates, demonstrating excellent agreement. The IA parameters $\mucen$ and $\musat$ do not affect galaxy number counts and have zero gradient everywhere. These gradients enable efficient gradient-based inference of HOD parameters from galaxy clustering observations.}
    \label{fig:hodgradients}
\end{figure*}

We first study gradients of the 1-pt statistics of the galaxy catalog, $\langle\ncen\rangle$ and $\langle\nsat\rangle$, for various halo masses evaluated at the fiducial HOD values.
We compare the autodifferentiation (autodiff) values from \diffhodia with finite-difference methods.
The IA parameters, $\mucen$ and $\musat$, are excluded in this analysis as they do not affect the galaxy number counts.
The results of this analysis are shown in Figure \ref{fig:hodgradients}.

We see in the left panel of Figure \ref{fig:hodgradients} that the only HOD parameters with non-zero gradients for $\langle\ncen\rangle$ are $\log M_\text{min}$ and $\sigma_{\log M}$, which is in agreement with the analytic expression for $\langle\ncen\rangle$ given in Equation \ref{eqn:ncen}. 
The finite difference gradients are additionally plotted as scatter, exhibiting excellent agreement with the autodiff gradients.
The gradients vanish for $\log_{10}M/M_\odot \gtrsim 13$, at which point all available halos in the catalog are populated with a central galaxy.
Similarly, at $\log_{10}M/M_\odot \lesssim 11$, the halos are not massive enough to frequently host central galaxies, so the gradients become small.
The gradients for both $\sigma_{\log M}$ and $\log M_\text{min}$ are approximately unity (up to a sign) at $\log_{10}M/M_\odot \approx 12$, where the occupation probability for the fiducial HOD transitions from 0 to 1.

In the right panel of Figure \ref{fig:hodgradients}, we see that all five HOD parameters have nonzero gradients for $\langle\nsat\rangle$, as expected from the analytic expression given in Equation \ref{eqn:nsat}.
The magnitudes of the gradients vary across several orders of magnitude as halo mass increases, which is an important diagnostic for specifying learning rates when using the gradients in a gradient-based optimization pipeline.
We again see excellent agreement between the \diffhodia autodiff gradients and the finite differences.
These results are in excellent agreement with a similar analysis shown in \citet{horowitz2022}.

\subsection{Intrinsic Alignment Gradients in \diffhodia}

\begin{figure*}
    \centering
    \includegraphics[width=0.9\textwidth]{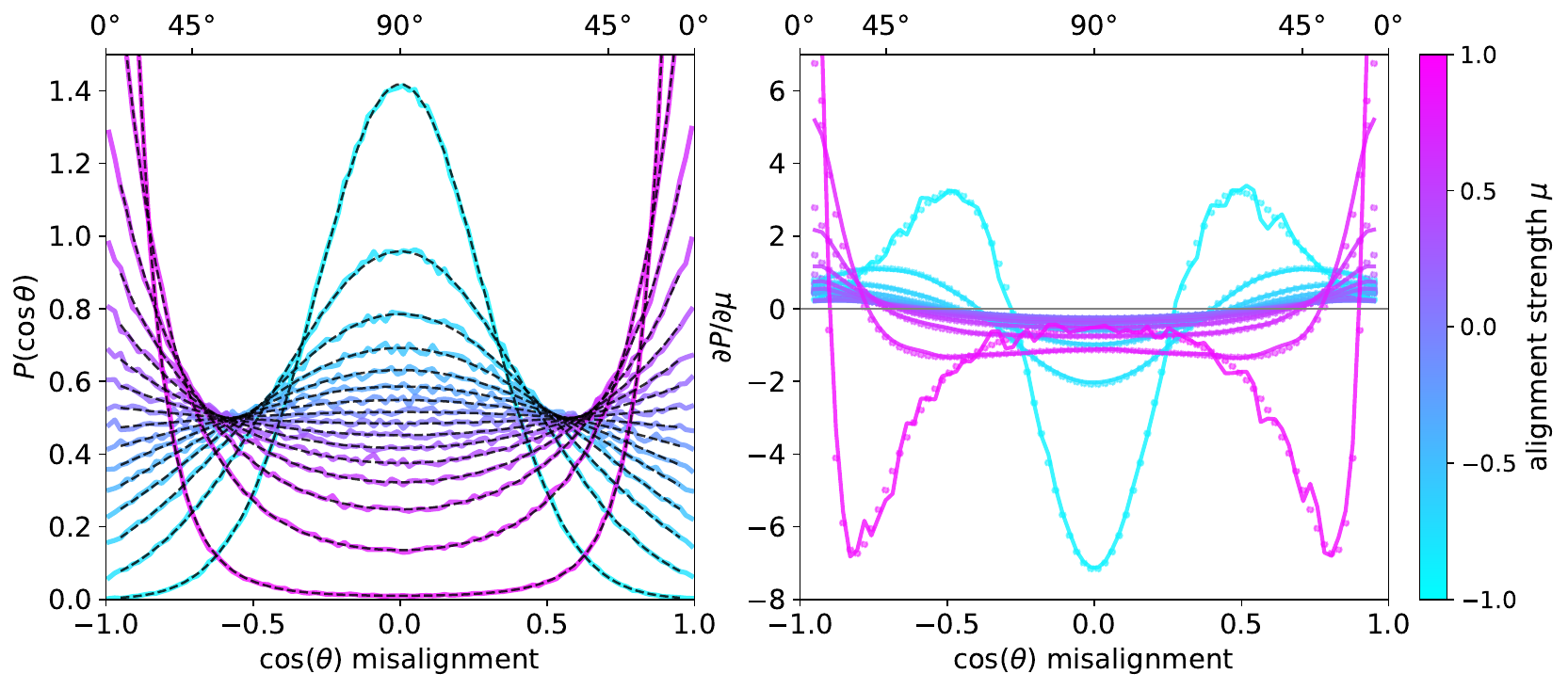}
    \caption{Validation of differentiable sampling from the Dimroth--Watson distribution for galaxy-halo misalignment angles. \textbf{Left panel:} Probability distribution $P(\cos\theta)$ of misalignment angles for varying alignment strength $\mu$. 
    The top axes show the corresponding misalignment angle $\theta$ in degrees.
    Solid lines show histograms from samples drawn using our differentiable inverse-CDF sampler; dashed lines show the analytic Dimroth--Watson PDF. The close agreement validates the differentiable sampling implementation. Positive $\mu$ (magenta) produces alignment with probability concentrated at $\cos\theta = \pm 1$, while negative $\mu$ (cyan) produces anti-alignment peaked at $\cos\theta = 0$. \textbf{Right panel:} Gradient of the probability distribution with respect to the alignment parameter, $\partial P / \partial \mu$. Dashed lines show analytic gradients derived from the PDF formula; scatter points show finite-difference gradients of the analytic Dimroth--Watson PDF; solid lines show gradients computed via automatic differentiation through the sampling procedure, where a Gaussian kernel density estimate is used to obtain a smooth density from discrete samples. 
    }
    \label{fig:ia_gradients}
\end{figure*}

We next examine the differentiability of the IA model in \diffhodia by validating the accuracy and gradients of the Dimroth--Watson distribution sampling procedure.
The Dimroth--Watson distribution is parameterized by the alignment strength $\mu \in [-1, 1]$, which controls the shape of the Dimroth--Watson distribution from which galaxy misalignment angles are sampled.
We compare the differentiable Dimroth--Watson samples with typical samples for a range of misalignment angles in Figure \ref{fig:ia_gradients}.
The \diffhodia data was generated using $500000$ samples and 6 Newton iterations for the inverse CDF sampling procedure.
For each sample, we compute the misalignment angle $\theta$ between the sampled galaxy orientation and a reference alignment direction.

The left panel of Figure \ref{fig:ia_gradients} shows the empirical probability distributions $P(\cos\theta)$ constructed from the samples (solid lines) alongside the analytic Dimroth--Watson PDF (dashed lines).
The histograms are computed with 100 bins spanning $\cos\theta \in [-1, 1]$ and normalized to unit area.
The close agreement across all values of $\mu$ validates our implementation of the inverse-CDF sampling procedure for sampling from the Dimroth--Watson.
As expected, positive $\mu$ produces alignment with probability mass concentrated near $\cos\theta = \pm 1$ (corresponding to $\theta \approx 0^\circ$ or $180^\circ$), while negative $\mu$ produces anti-alignment with probability peaked at $\cos\theta = 0$ ($\theta \approx 90^\circ$).

To validate the gradients, we compute $\partial P(\cos\theta) / \partial \mu$ in three ways.
First, we use autodiff through the \diffhodia sampling procedure.
We draw samples for each $\mu$ value, construct a differentiable Gaussian kernel density estimate (KDE) to obtain a smooth density function from the discrete samples, and compute gradients via autodiff.
The KDE is necessary to convert the discrete histogram into a continuous, differentiable function. 
We additionally compute analytic and finite-difference gradients of the exact Dimroth--Watson PDF with respect to $\mu$, which can be derived from Equation \ref{eqn:dimrothwatson} for the analytic case.
The right panel of Figure \ref{fig:ia_gradients} shows generally good agreement between the autodiff gradients (solid lines), the analytic gradients (dashed lines), and finite-difference gradients (scatter).
We note that in the case of $\mu \approx 0$, the gradient similarly tends to zero as the Dimroth--Watson distribution becomes uniform.
The gradient can thus be sensitive to Monte Carlo noise in this regime.

\section{Applications of differentiable capabilities}
\label{sec:experiments}

We demonstrate the utility of \diffhodia with gradient-based optimization experiments.
These experiments illustrate how gradients in the simulation can be used to obtain parameter estimates and posteriors over parameters of interest from mock observational data.
All experimental results presented use a value of $N_{\text{max}} = 48$, $\tau = 0.1$, and \texttt{SubhaloPhaseSpace} with a \texttt{NFWPhaseSpace} fallback.
A constant radial alignment strength model is used for the subhalo IA.
This combination of phase space and alignment model is sufficient to accurately model the IA of \textsc{tng300}, as studied in \citet{vanalfen_2023}.
As our construction follows \citet{horowitz2022} closely, and having validated both the accuracy and differentiability of \diffhodia against \halotoolsia, our emphasis will be on the IA parameters $\mucen$ and $\musat$;
several experiments illustrating the differentiability of the HOD are included in \citet{horowitz2022}.
We construct moment-matching and differentiable correlation function optimization objectives.
In addition, we demonstrate accelerated inference with \diffhodia using HMC, compared to \halotoolsia with MCMC.

\subsection{Moment-matching Objective}
\label{sec:momentmatching}

\textbf{Setup.} We consider the task of inferring the IA parameters $\boldsymbol{\mu} = (\mucen, \musat)$ from a target galaxy catalog observable, restricting for simplicity to a single HOD configuration given by the fiducial \textsc{tng300} model used previously.
Importantly, fixing the HOD does not fix the realized galaxy catalog: different random seeds produce catalogs with varying numbers of galaxies, requiring an optimization procedure that is robust to stochastic catalog realizations.
To this end, we consider a \textit{moment matching} optimization procedure, matching the misalignment angle distributions between the generated and target galaxy catalogs via their first two moments.
This does not rely on a one-to-one correspondence between galaxies, and additionally requires no knowledge of the underlying PDF governing the galaxy misalignments.

\textbf{Optimization.} For a given parameter vector $\boldsymbol{\mu}$ and simulation seed $s$, we generate the mock galaxy catalog and compute the per-galaxy alignment statistic $t^2 = (\mathbf{n} \cdot \mathbf{u})^2$, where $\mathbf{n}$ is the galaxy orientation and $\mathbf{u}$ is the reference axis (host halo major axis for centrals, radial direction for satellites).
We use $t^2$ rather than $t$ because the Dimroth--Watson distribution is symmetric about $t=0$, making $\langle t \rangle = 0$ for all $\mu$.
For each population (centrals and satellites separately), we compute the mean $\langle t^2 \rangle$ and variance $\mathrm{Var}(t^2)$ across all galaxies, which serve as our summary statistics.

Our loss function matches these first and second moments between simulated and observed catalogs across seeds $s$, treating central and satellite galaxy populations separately 
\begin{equation}
\label{eqn:loss}
\mathcal{L}(\boldsymbol{\mu}; s) = \alpha_{\text{cen}} \mathcal{L}_{\text{cen}} + \mathcal{L}_{\text{sat}},
\end{equation}
where
\begin{align}
\label{eqn:lcenlsat}
\mathcal{L}_{\text{cen}} &=  \left(\langle t^2 \rangle_{\mathrm{sim}} - \langle t^2 \rangle_{\mathrm{obs}} \right)^2 \nonumber \\
&\quad + w_{\text{cen},\sigma^2} \left(\mathrm{Var}[t^2]_{\mathrm{sim}} - \mathrm{Var}[t^2]_{\mathrm{obs}} \right)^2, \\
\mathcal{L}_{\text{sat}} &=  \left(\langle t^2 \rangle_{\mathrm{sim}} - \langle t^2 \rangle_{\mathrm{obs}} \right)^2 \nonumber \\
&\quad + w_{\text{sat},\sigma^2} \left(\mathrm{Var}[t^2]_{\mathrm{sim}} - \mathrm{Var}[t^2]_{\mathrm{obs}} \right)^2,
\end{align}
where it is implied that the $t^2$ statistics are computed over central galaxies only in $\mathcal{L}_\text{cen}$ and satellites only in $\mathcal{L}_{\text{sat}}$.
We use weights $w_{\text{cen},\sigma^2} = w_{\text{sat},\sigma^2} = 0.5$, and $\alpha_{\text{cen}} = 2.0$ to up-weight the smaller central galaxy population.
To mitigate the Monte Carlo noise in Dimroth--Watson sampling, we average the loss over $N_s = 3$ random seeds:
\begin{equation}
\bar{\mathcal{L}}(\boldsymbol{\mu}) = \frac{1}{N_s} \sum_{i=1}^{N_s} \mathcal{L}(\boldsymbol{\mu}; s_i).
\end{equation}
This construction matches the variance $\mathrm{Var}[t^2]$ within each catalog to capture galaxy shape noise, while averaging the loss over multiple simulated realizations stabilizes the gradient signal.
The seed averaging is performed over a fixed set of seeds; importantly, the target catalog uses a different seed than the optimization seeds.
We compare experiments using one and three seeds in Figure~\ref{fig:gradient}.
More detail on optimization hyperparameters is given in Appendix \ref{app:hyperparams}.

\begin{figure}
    \centering
    \includegraphics[width=\linewidth]{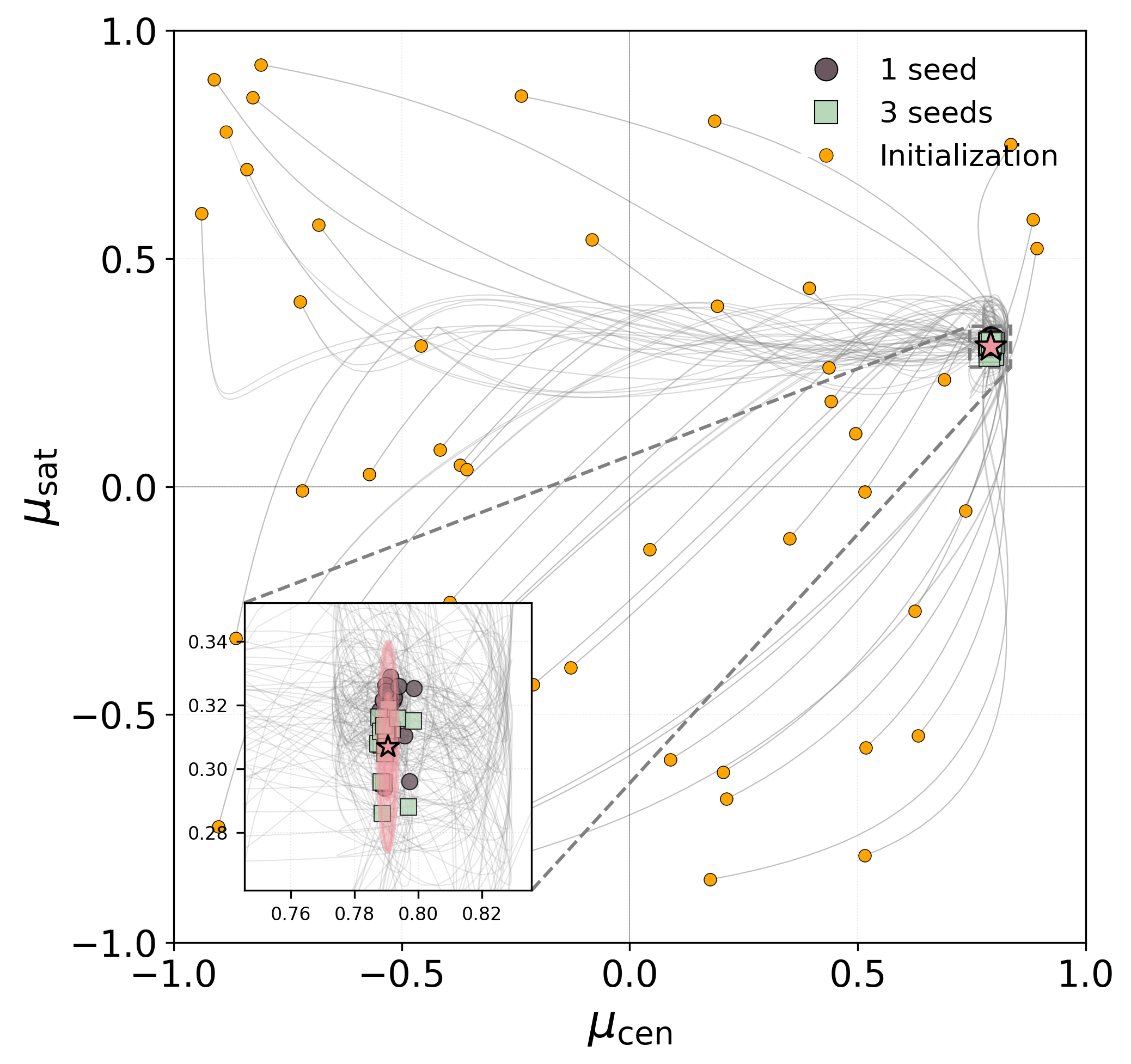}
    \caption{Gradient-based recovery of IA parameters from 50 random initializations using moment-matching optimization. The target parameters ($\mu_{\rm cen} = 0.7905$, $\mu_{\rm sat} = 0.307$, pink star) represent the best-fit $\mu$ values to the \textsc{tng300} HOD configuration, with uncertainty contours in pink estimated from the scatter in effective $\mu$ values recovered by matching $\langle t^2 \rangle$ to the theoretical Dimroth--Watson expectation across 50 independent HOD realizations. Optimization trajectories are shown as gray lines connecting initial positions to converged solutions. Gray circles denote optimizations using a single HOD realization seed per gradient step, while green squares show results when averaging over three seeds. The inset panel (lower left) shows a zoomed view of the convergence region, revealing tight clustering of final parameter estimates around the true values. Both single-seed and three-seed strategies successfully recover the target parameters across diverse initializations.}
    \label{fig:gradient}
\end{figure}

We note that for the target catalog generated at fixed input parameters, the observed alignment angles represent a single Monte Carlo draw from the underlying Dimroth--Watson distribution, inducing an effective $\boldsymbol{\mu}$ that may differ slightly from the nominal input values.
To determine the effective $\boldsymbol{\mu}$ values for the target catalog, we independently match the observed $\langle t^2 \rangle$ to the theoretical Dimroth--Watson expectation for the central and satellite populations.
This yields best-fit values of $\mucen = 0.7905$ and $\musat = 0.307$, which differ slightly from the input values $\mucen = 0.79$ and $\musat = 0.30$.

We estimate the uncertainties in $\mucen$ and $\musat$ by quantifying the variance in effective $\mu$ values across independent catalog realizations.
To this end, we generate 50 independent HOD realizations at the fiducial IA parameters using different random seeds.
For each realization, we compute $\langle t^2 \rangle$ separately for centrals and satellites, as done in the optimization, and determine the effective $\mu$ by matching to the theoretical Dimroth--Watson expectation via numerical inversion.
The scatter in these effective $\mu$ values across realizations quantifies the uncertainty in the effective $\mu$ for a single catalog realization, capturing the combined variance from HOD sampling and orientation sampling.
Importantly, this also defines the expected region within which our optimization should converge.
This yields uncertainties $\sigma(\mucen) = 0.001$ and $\sigma(\musat) = 0.012$, with a correlation coefficient of $\rho = 0.04$.
These uncertainties are visualized in pink in Figure~\ref{fig:gradient}.

We note that the roughly one order of magnitude larger uncertainty for $\musat$ compared to $\mucen$ is primarily due to the smaller alignment strength of $\musat=0.30$ relative to $\mucen=0.79$
At lower $\mu$, the Dimroth--Watson distribution is less sharply peaked and galaxy orientations are more dispersed, producing greater variance in $\mu$ inferred from a finite catalog.
This larger uncertainty is reflected in the scatter of converged $\musat$ values seen in Figure~\ref{fig:gradient}.

\textbf{Results.} Figure~\ref{fig:gradient} shows the results of gradient-based optimization of IA parameters using the moment matching objective.
We perform 50 independent optimization runs from random initializations uniformly sampled from $\mu \in [-1.0, 1.0]$, each run for 100 optimization steps.
The trajectories in Figure \ref{fig:gradient} show the parameter space flow from initializations to the fiducial \textsc{tng300} IA values, with final points filtered by best loss.
We see that the moment-matching loss over the full galaxy field successfully optimizes the IA parameters toward the ground truth values for the HOD across all initializations and both seed-averaging strategies.
In the inset panel, we observe that the 3-seed optimization exhibits slightly tighter convergence to the true values.
Quantitatively, the 1-seed optimization results in final values $\mucen = 0.791 \pm 0.002$ and $\musat = 0.320 \pm 0.007$, while the 3-seed optimization achieves $\mucen = 0.790 \pm 0.002$ and $\musat = 0.309 \pm 0.006$.

The scatter in converged $\mucen$ values is slightly larger than the computed target uncertainty, likely reflecting optimization noise and finite optimization time.
Conversely, the scatter in converged $\musat$ values is comparable to the target uncertainty, and the slight bias toward higher values may arise from an asymmetry of the loss landscape near the true value.
Nevertheless, all optimization runs converge to within a few percent of the true values, demonstrating robust parameter recovery.
Although the IA parameters in this simplified setting could be recovered via a direct curve fit, this experiment illustrates how differentiability enables efficient parameter recovery using only one-point alignment statistics.
This is especially relevant when the analytic form of the PDF is not known; in this case, we have captured properties of the true distribution as characterized by its first two moments.

\subsection{Correlation Function Objective}
\label{sec:correlationfunctionobjective}

\begin{figure}
    \centering
    \includegraphics[width=\linewidth]{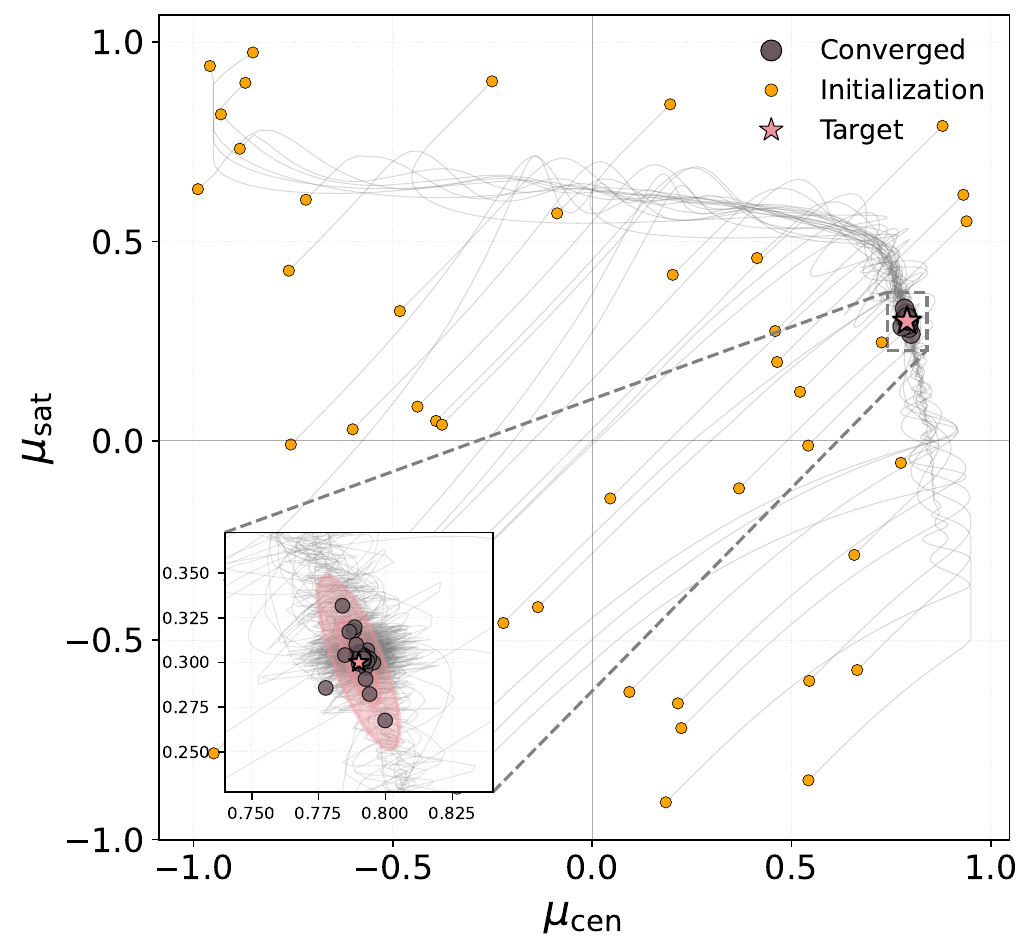}
    \caption{Gradient-based recovery of IA parameters from 50 random initializations using correlation function matching optimization. The target parameters ($\mu_{\rm cen} = 0.79$, $\mu_{\rm sat} = 0.30$, pink star) represent the fiducial \textsc{tng300} HOD configuration, with $1\sigma$ and $2\sigma$ uncertainty regions estimated from the scatter of maximum-likelihood fits across 50 independent HOD realizations shown in pink. Optimization trajectories are shown as gray lines connecting initial positions to converged solutions. Gray circles denote converged values. The inset panel (lower left) shows a zoomed view of the convergence region, revealing tight clustering of final parameter estimates around the true values.}
    \label{fig:omega_optimization}
\end{figure}

\textbf{Setup}. We now turn to optimization with a more traditional summary statistic with 2PCFs.
In this section, we will optimize the IA parameters such that the generated catalog $\omega(r)$ correlation matches that of the fiducial \textsc{tng300} correlation function, denoted $\hat{\omega}(r)$.
$\hat{\omega}(r)$ is computed by averaging over 20 independent orientation samples at the fiducial IA parameters.
The per-bin variance of $\hat{\omega}(r)$ is estimated from the scatter across these realizations, which reduces galaxy shape noise in the optimization target.
Unlike a single realization of galaxy orientations, which can result in effective $\boldsymbol{\mu}$ values that differ from the input, this averaging ensures that the optimal $\boldsymbol{\mu}$ parameters for $\hat{\omega}(r)$ converge to the true input values.
We choose to optimize on this statistic over $\eta(r)$, as $\omega(r)$ is less contaminated with galaxy shape noise due to it being a cross correlation with galaxy positions.
We cannot use $\xi(r)$ as it has no dependence on $\boldsymbol{\mu}$; a full analysis that includes the HOD parameters would jointly use $\xi(r)$, $\omega(r)$, and $\eta(r)$.

\textbf{Optimization.} We construct a weighted mean-squared-error loss between the predicted and target correlation functions:
\begin{equation}
\mathcal{L}(\boldsymbol{\mu}) = \sum_{k} W_k \left[\omega(r_k) - \hat{\omega}(r_k)\right]^2,
\end{equation}
where $\hat{\omega}(r_k)$ is the averaged target correlation computed from the fiducial \textsc{tng300} catalog and $W_k$ are inverse-variance weights, determined by estimating the variance of $\hat{\omega}(r_k)$ across multiple orientation realizations at fixed positions.
We generate $20$ independent orientation samples at reference IA parameter values and compute:
\begin{equation}
W_k = \frac{1}{\sigma^2_{\omega}(r_k) + \epsilon},
\end{equation}
where $\sigma^2_{\omega}(r_k)$ is the empirical variance in bin $k$ and $\epsilon = 10^{-6}$ provides numerical stability to prevent division by zero in bins with very low variance.
The weights are normalized such that $\sum_k W_k = 1$.
These inverse-variance weights serve only to improve convergence by preventing high-variance bins from dominating the gradient updates, and approximates the covariance due to galaxy shape noise, which is in general dominant over the sample variance at the scales being considered \citep{vanalfen_2023}.
While the jackknife covariances that account for both sample variance and galaxy shape noise from the \textsc{TNG300} data itself are available following the procedure of \citet{vanalfen_2023}, we instead use a diagonal variance estimate obtained by averaging over many \diffhodia orientation realizations (for fixed galaxy positions), which serves as a proxy for the shape noise and is considerably more computationally efficient.
More detail on optimization hyperparameters is given in Appendix \ref{app:hyperparams}.

\textbf{Results.} Figure~\ref{fig:omega_optimization} shows the results of gradient-based optimization of IA parameters using the differentiable $\omega(r)$ objective.
We perform 50 independent optimization runs from random initializations uniformly sampled from $\mu \in [-1.0, 1.0]$, each run for 2000 optimization steps.
The optimization successfully recovers the target parameters across all initializations, with mean recovered values $\mucen = 0.791 \pm 0.003$ and $\musat = 0.303 \pm 0.009$, compared to target values of $\mucen = 0.79$ and $\musat = 0.30$.
All converged solutions lie well within the expected statistical uncertainty, shown in pink in Figure \ref{fig:omega_optimization}.

To estimate the uncertainty, we generate 50 independent HOD catalog realizations at the true IA parameters with different random seeds.
For each realization, we compute $\omega(r)$ averaged over 20 orientation samples and find the maximum likelihood IA parameters via grid search, minimizing
\begin{equation}
\chi^2 = ({\omega} - \hat{{\omega}})^T \mathbf{C}^{-1} ({\omega} - \hat{{\omega}}),
\end{equation}
where $\mathbf{C}$ is the covariance matrix of $\omega(r)$ estimated from 50 independent orientation samples at the fiducial parameters.
We additionally correct for the finite number of realizations by incorporating the Hartlap factor \citep{Hartlap_2006} when inverting the covariance matrix.
We emphasize that this empirical covariance is distinct and more accurate than the diagonal variance estimate used in the optimization.

The scatter in recovered parameters across these 50 catalogs provides an empirical estimate of the covariance, yielding uncertainties $\sigma(\mucen) = 0.008$ and $\sigma(\musat) = 0.026$, with correlation coefficient $\rho = -0.68$.
This anti-correlation aligns with theoretical expectations and was also seen experimentally in \citet{berman2025softclustering}.
The tighter clustering of converged values reflects the fact that all optimization runs fit the same catalog to the same target, whereas the uncertainty contours capture the full variance including HOD stochasticity.

\subsection{Hamiltonian Monte Carlo}

\begin{figure}
    \centering
    \includegraphics[width=\columnwidth]{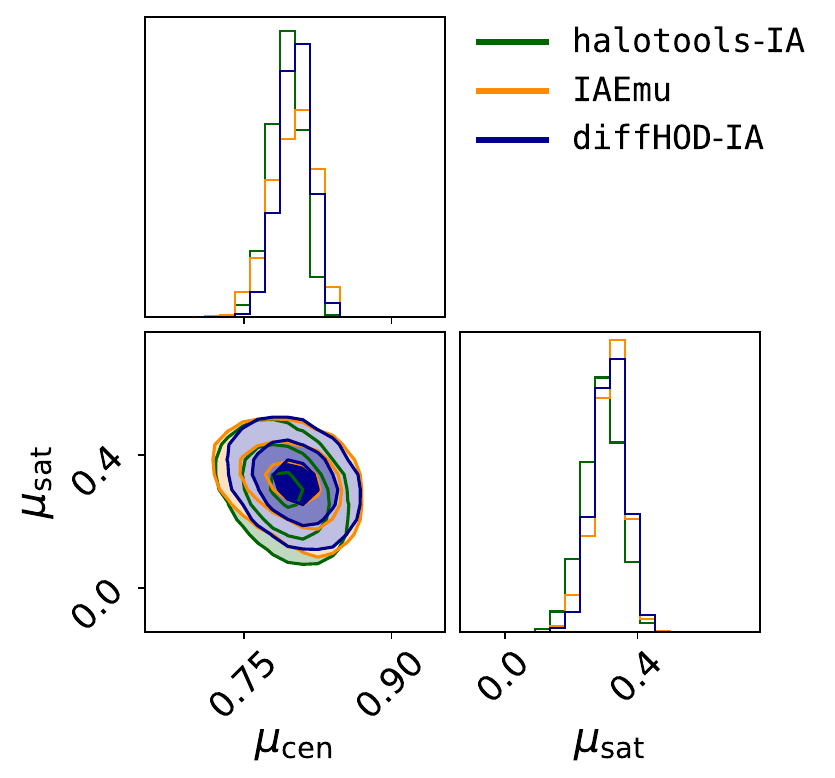}
    \caption{$\mucen$ and $\musat$ posteriors for the fiducial \textsc{tng300} catalog derived using \diffhodia with HMC (blue), \IAEmu with HMC (orange), and \halotoolsia with MCMC (green). \diffhodia is in excellent agreement with \halotoolsia, while exhibiting substantially faster inference. Exact posterior values are given in Table \ref{tab:hmc_comparison}.}
    \label{fig:posteriors}
\end{figure}

\textbf{Setup.} To complement our previous correlation function matching gradient descent experiment, we now turn to HMC to illustrate Monte Carlo inference capabilities with \diffhodia.
We use the same $\hat{\omega}(r)$ mock observation from Section \ref{sec:correlationfunctionobjective}; however, where a diagonal variance estimate sufficed for gradient-based optimization, here we adopt the jackknife covariance from \citet{Pandya_2025} specifically to enable a direct comparison of posteriors between \diffhodia, \halotoolsia, and \IAEmu, all of which used identical covariance estimates in that work.
For this analysis, we use a fixed HOD and only consider $\musat$ and $\mucen$ as free parameters.

We use the uninformative priors:
\begin{equation}
    \mucen, \musat \sim \text{Uniform}(-1,1),
\end{equation}
and employ the $\tt{numpyro}$ \citep{phan2019composable} implementation of a No-U-Turn Sampler (NUTS, \cite{hoffman2011nouturnsampleradaptivelysetting}).
We use four chains with 1500 steps (500 burn-in), and additionally average over 10 likelihood realizations per step to aid convergence.
A similar analysis was conducted in \citet{Pandya_2025} using \halotoolsia and \IAEmu, allowing us to compare \diffhodia against both the ground truth \halotoolsia and its emulator \IAEmu.
HMC on \diffhodia took roughly 5 minutes to converge on a single NVIDIA A100 GPU, while HMC with \IAEmu converged in approximately one minute on the same GPU. 
As \halotoolsia is not differentiable, a comparable MCMC analysis required up to a full day across 150 CPU cores.

\begin{table}
\centering
\caption{Comparison of $\mucen$ and $\musat$ posteriors from the Monte Carlo experiments. \diffhodia and \IAEmu posteriors used HMC, while \halotoolsia posteriors used MCMC. }
\label{tab:hmc_comparison}
\begin{tabular*}{\columnwidth}{@{\extracolsep{\fill}}lcc}
\hline
Method & $\mucen$ & $\musat$ \\
\hline
\halotoolsia & $0.793 \pm 0.017$ & $0.294 \pm 0.056$ \\
\IAEmu & $0.799 \pm 0.022$ & $0.317 \pm 0.049$ \\
\diffhodia & $0.802 \pm 0.016$ & $0.318 \pm 0.048$ \\
\hline
\end{tabular*}
\end{table}

\textbf{Results.} Posteriors for \diffhodia, \halotoolsia, and \IAEmu for this experiment are shown in Figure \ref{fig:posteriors}.
There is in general excellent overlap between the three posteriors, illustrating that \diffhodia offers similar inference capabilities as \halotoolsia with much faster convergence with HMC.
A slight bias is seen in the case of $\musat$ for both \diffhodia and \IAEmu compared to \halotoolsia. This can potentially be attributed to different seeds in the target galaxy catalog.
Nonetheless, \diffhodia is within $0.4\sigma$ agreement with \halotoolsia, which is consistent with the expected variation for catalogs generated with different random seeds.
Exact posterior values are given in Table \ref{tab:hmc_comparison}.

\section{Summary \& Discussion}
\label{sec:summary}

In this work, we have developed \diffhodia, a differentiable HOD framework that includes galaxy IA modeling.
Our HOD implementation closely follows that of \citet{horowitz2022}, which is extended to the IA modeling of \citet{vanalfen_2023} via inverse Dimroth--Watson CDF sampling.
We additionally draw inspiration from \citet{Hearin_2022} to extend the differentiability of \diffhodia to include IA correlation functions.
% \jab{Should move the corr funct stuff up to main body.}
The \diffhodia code is publicly available at \githubmaster\footnote{https://github.com/snehjp2/diffHOD-IA}.

\textbf{Speed.} The utility of \diffhodia over \halotoolsia stems from its differentiability, which enables the use of gradient-based algorithms to optimize the input model parameters.
For forward modeling, \diffhodia runs in comparable time to \halotoolsia (seconds per catalog on CPU), with no significant speedup on GPU.
For forward modeling, \IAEmu is substantially faster, exhibiting an approximately $10000 \times$ speedup over \halotoolsia and \diffhodia.
The benefits of differentiability are clearer in inverse modeling. 
Our HMC analysis converged in approximately 5 minutes on a single GPU, compared to approximately one day across 150 CPU cores for \halotoolsia with MCMC.

The utility of \diffhodia over \IAEmu comes from its differentiability at the catalog level.
\IAEmu models the correlations $\xi(r)$, $\omega(r)$, and $\eta(r)$, bypassing the galaxy catalog generation step.
This limits its predictive abilities to only 2PCFs, whereas \diffhodia can be extended to differentiably model any summary statistic.
This also requires that any changes to the HOD formulation require retraining \IAEmu, while these changes, if necessary, can be made directly within the \diffhodia simulation.
In addition, \IAEmu may not generalize well to different dark matter catalogs (e.g., different cosmologies), whereas this is not a concern with \diffhodia.

\diffhodia is written in JAX \citep{jax2018github}, enabling several processing and vectorization speedups via \texttt{jax.jit} and \texttt{jax.vmap}.
As vectorization on GPU requires static array sizes, modeling up to the 2PCF level can be easily vectorized.
This does not include generating galaxy catalogs, as the sizes of galaxy catalogs vary across HOD instances and random seeds.
For potentially different galaxy-halo connections, or with using weighted-galaxy catalogs, static galaxy catalogs can be generated in parallel.

\textbf{Results.}~We benchmarked the accuracy of the sampling procedure and gradients of \diffhodia, comparing autodiff computed gradients with analytic and finite-differences, and comparing the differentiable Dimroth--Watson samples with samples from the true distribution.
\diffhodia agreed with \halotoolsia across all tests, including galaxy number counts, and the $\xi(r)$, $\omega(r)$, and $\eta(r)$ statistics for the fiducial galaxy catalog.

For a fixed HOD, we utilized gradients in \diffhodia to retrieve IA parameters $\boldsymbol{\mu}$ from a mock observable galaxy catalog and correlations corresponding to the \textsc{tng300} simulation.
We tested this using gradient-based optimization with a moment matching and correlation function loss.
We further leveraged the differentiability of \diffhodia in a Hamiltonian Monte Carlo pipeline, showing excellent agreement with \halotoolsia and \IAEmu.

% \textbf{Limitations.}~There are several potential shortcomings of this implementation.
% First, the Binomial approximation to the Poisson distribution for satellite occupation can introduce a slight variance mismatch for finite $N_\text{max}$.
% Second, the relaxed Bernoulli sampling with temperature $\tau = 0.1$ introduces smoothing that manifests as slightly larger variance in galaxy number counts compared to \halotoolsia. 
% Third, we observed numerical instabilities in the autodiff gradients of the Dimroth--Watson distribution near $\mu \approx 0$, where the gradient signal is weak and dominated by sampling noise.
% These issues do not significantly impact the inference results presented, but should be considered when applying \diffhodia.

\textbf{Future Work.} Future extensions of this work could proceed in several directions. The current implementation uses the \citet{Zheng_2007} HOD formulation, but the differentiable framework readily accommodates more sophisticated HOD models that include assembly bias \citep{Hearin2016} or environment-dependent effects \citep{Yuan_2018}.
In addition, \halotoolsia specifies galaxy misalignments solely according to galaxy orientations, and does not currently include galaxy shapes or ellipticities.
\diffhodia can readily be extended with shape information in accordance with future versions of \halotoolsia.
Similarly, while we have focused on the radial alignment model with constant alignment strength, the distance-dependent alignment strength model implemented in \diffhodia could be explored for galaxies with radially-varying alignment properties.
More generally, this can be extended to different IA parameters \emph{per galaxy}, as opposed to catalog-wide definitions.

The differentiability also allows inserting \diffhodia into larger differentiable inference pipelines.
For example, \diffhodia can be integrated into simulation pipelines that incorporate a differentiable particle mesh solver such as \textsc{jaxPM} \citep{li2022pmwddifferentiablecosmologicalparticlemesh}, along with a differentiable halo finder like \textsc{jFoF} \citep{horowitz2025jfofgpuclusterfinding}.
This would enable end-to-end gradient flow from cosmological, HOD, and IA parameters to the galaxy field.

The differentiable correlation function framework could also be extended to 2D-projected statistics, which are more directly comparable to observational weak lensing measurements \citep{vanalfen_2025_inprep}.
Of equal interest is the extension to higher-order statistics and field level inference methods which would impose tighter constraints on parameters.
A joint inference over both HOD and IA parameters represents a natural application, using the differentiable estimators for $\xi(r)$ and $\omega(r)$.
In the present work, fixing the HOD avoids degeneracies between HOD and IA parameters that arise at the 2PCF level --- for instance, the satellite fraction governed by $\log M_1$ and $\alpha$ directly affects the relative contributions of centrals and satellites to $\omega(r)$, which could partially mimic changes in $\mu_{\rm sat}$.
Such joint analyses over the full 7-dimensional parameter space would require careful treatment of these degeneracies and are reserved for future work.
These developments provide a foundation for efficient, gradient-based joint inference of HOD and IA parameters in current and future weak lensing surveys.

\section*{Acknowledgments}
We thank Nicholas Van Alfen, Alyssa Cordero, Hamza Ahmed, Robin Walters, Andrew Hearin, and Benjamin Horowitz for useful conversations in developing this work.
The computations in this paper were run on the FASRC Cannon cluster supported by the FAS Division of Science Research Computing Group at Harvard University.
This work is supported by NASA through a Roman Research and Support Participation program (NASA grant 80NSSC24K0088) and the OpenUniverse effort (JPL Contract Task 70-711320); by the DOE (DE-SC0024787); and by the NSF (AST2206563, AST2442796).

% \newpage
\bibliographystyle{mnras}
\bibliography{bib-3}

@ARTICLE{Siegel2025,
       author = {{Siegel}, J. and {McCullough}, J. and {Amon}, A. and {Lamman}, C. and
         {Jeffrey}, N. and {Joachimi}, B. and {Hoekstra}, H. and {Heydenreich}, S. and
         {Ross}, A.~J. and {Aguilar}, J. and {Ahlen}, S. and {Bianchi}, D. and
         {Blake}, C. and {Brooks}, D. and {Castander}, F.~J. and {Claybaugh}, T. and
         {de la Macorra}, A. and {DeRose}, J. and {Doel}, P. and {Emas}, N. and
         {Ferraro}, S. and {Font-Ribera}, A. and {Forero-Romero}, J.~E. and
         {Gazta{\~n}aga}, E. and {Gontcho A Gontcho}, S. and {Gutierrez}, G. and
         {Honscheid}, K. and {Ishak}, M. and {Joudaki}, S. and {Kehoe}, R. and
         {Kirkby}, D. and {Kisner}, T. and {Krolewski}, A. and {Lahav}, O. and
         {Lambert}, A. and {Landriau}, M. and {Le Guillou}, L. and {Levi}, M.~E. and
         {Manera}, M. and {Meisner}, A. and {Miquel}, R. and {Moustakas}, J. and
         {Nadathur}, S. and {Newman}, J.~A. and {Niz}, G. and
         {Palanque-Delabrouille}, N. and {Percival}, W.~J. and {Porredon}, A. and
         {Prada}, F. and {P{\'e}rez-R{\`a}fols}, I. and {Rossi}, G. and
         {Sanchez}, E. and {Saulder}, C. and {Schlegel}, D. and {Schubnell}, M. and
         {Semenaite}, A. and {Silber}, J. and {Sprayberry}, D. and {Sun}, Z. and
         {Tarl{\'e}}, G. and {Weaver}, B.~A. and {Zhou}, R. and {Zou}, H.},
        title = "{Intrinsic alignment demographics for next-generation lensing: Revealing 
                  galaxy property trends with DESI Y1 direct measurements}",
      journal = {arXiv e-prints},
         year = 2025,
        month = jul,
          eid = {arXiv:2507.11530},
        pages = {arXiv:2507.11530},
archivePrefix = {arXiv},
       eprint = {2507.11530},
 primaryClass = {astro-ph.CO},
}

@ARTICLE{NavarroGirones2025,
       author = {{Navarro-Gironés}, D. and {Crocce}, M. and {Gazta{\~n}aga}, E. and
         {Wittje}, A. and {Siudek}, M. and {Hoekstra}, H. and {Hildebrandt}, H. and
         {Joachimi}, B. and {Paviot}, R. and {Baugh}, C.~M. and {Carretero}, J. and
         {Casas}, R. and {Castander}, F.~J. and {Eriksen}, M. and {Fernandez}, E. and
         {Fosalba}, P. and {Garc{\'i}a-Bellido}, J. and {Miquel}, R. and
         {Padilla}, C. and {Renard}, P. and {S{\'a}nchez}, E. and {Serrano}, S. and
         {Sevilla-Noarbe}, I. and {Tallada-Crespí}, P.},
        title = "{The PAU Survey: Measuring intrinsic galaxy alignments in deep wide fields 
                  as a function of colour, luminosity, stellar mass and redshift}",
      journal = {arXiv e-prints},
         year = 2025,
        month = may,
          eid = {arXiv:2505.15470},
        pages = {arXiv:2505.15470},
archivePrefix = {arXiv},
       eprint = {2505.15470},
 primaryClass = {astro-ph.CO},
}

@ARTICLE{Georgiou2025,
       author = {{Georgiou}, Christos and {Chisari}, Nora Elisa and {Bilicki}, Maciej and
         {La Barbera}, Francesco and {Napolitano}, Nicola R. and {Roy}, Nivya and
         {Tortora}, Crescenzo},
        title = "{Intrinsic galaxy alignments in the KiDS-1000 bright sample: dependence 
                  on colour, luminosity, morphology, and galaxy scale}",
      journal = {arXiv e-prints},
         year = 2025,
        month = feb,
          eid = {arXiv:2502.09452},
        pages = {arXiv:2502.09452},
archivePrefix = {arXiv},
       eprint = {2502.09452},
 primaryClass = {astro-ph.CO},
}

@ARTICLE{Fortuna2025,
       author = {{Fortuna}, Maria Cristina and {Dvornik}, Andrej and {Hoekstra}, Henk and
         {Joachimi}, Benjamin and {Georgiou}, Christos and {Giblin}, Benjamin and
         {Heymans}, Catherine and {Hildebrandt}, Hendrik and {Kannawadi}, Arun and
         {Kuijken}, Konrad and {Wright}, Angus H.},
        title = "{KiDS-1000: Weak lensing and intrinsic alignment around luminous red 
                  galaxies}",
      journal = {arXiv e-prints},
         year = 2024,
        month = sep,
          eid = {arXiv:2409.15416},
        pages = {arXiv:2409.15416},
archivePrefix = {arXiv},
       eprint = {2409.15416},
 primaryClass = {astro-ph.CO},
}

@ARTICLE{Samuroff2023,
       author = {{Samuroff}, S. and {Mandelbaum}, R. and {Blazek}, J. and {Campos}, A. and
         {MacCrann}, N. and {Zacharegkas}, G. and {Amon}, A. and {Prat}, J. and
         {Singh}, S. and {Elvin-Poole}, J. and {Ross}, A.~J. and {Alarcon}, A. and
         {Baxter}, E. and {Bechtol}, K. and {Becker}, M.~R. and {Bernstein}, G.~M. and
         {Carnero Rosell}, A. and {Carrasco Kind}, M. and {Cawthon}, R. and
         {Chang}, C. and {Chen}, R. and {Choi}, A. and {Crocce}, M. and {Davis}, C. and
         {DeRose}, J. and {Dodelson}, S. and {Doux}, C. and {Drlica-Wagner}, A. and
         {Eckert}, K. and {Everett}, S. and {Fert{\'e}}, A. and {Gatti}, M. and
         {Giannini}, G. and {Gruen}, D. and {Gruendl}, R.~A. and {Harrison}, I. and
         {Herner}, K. and {Huff}, E.~M. and {Jarvis}, M. and {Kuropatkin}, N. and
         {Leget}, P. -F. and {Lemos}, P. and {McCullough}, J. and {Myles}, J. and
         {Navarro-Alsina}, A. and {Pandey}, S. and {Porredon}, A. and {Raveri}, M. and
         {Rodriguez-Monroy}, M. and {Rollins}, R.~P. and {Roodman}, A. and {Rossi}, G. and
         {Rykoff}, E.~S. and {S{\'a}nchez}, C. and {Secco}, L.~F. and
         {Sevilla-Noarbe}, I. and {Sheldon}, E. and {Shin}, T. and {Troxel}, M.~A. and
         {Tutusaus}, I. and {Weaverdyck}, N. and {Yanny}, B. and {Yin}, B. and
         {Zhang}, Y. and {Zuntz}, J.},
        title = "{The Dark Energy Survey Year 3 and eBOSS: constraining galaxy intrinsic 
                  alignments across luminosity and colour space}",
      journal = {\mnras},
         year = 2023,
        month = sep,
       volume = {524},
       number = {2},
        pages = {2195--2223},
          doi = {10.1093/mnras/stad2013},
archivePrefix = {arXiv},
       eprint = {2212.11319},
 primaryClass = {astro-ph.CO},
}

@ARTICLE{Johnston2019,
       author = {{Johnston}, Harry and {Georgiou}, Christos and {Joachimi}, Benjamin and
         {Hoekstra}, Henk and {Chisari}, Nora Elisa and {Farrow}, Daniel and
         {Fortuna}, Maria Cristina and {Heymans}, Catherine and {Joudaki}, Shahab and
         {Kuijken}, Konrad and {Wright}, Angus},
        title = "{KiDS+GAMA: Intrinsic alignment model constraints for current and future 
                  weak lensing cosmology}",
      journal = {\aap},
         year = 2019,
        month = apr,
       volume = {624},
          eid = {A30},
        pages = {A30},
          doi = {10.1051/0004-6361/201834714},
archivePrefix = {arXiv},
       eprint = {1811.09598},
 primaryClass = {astro-ph.CO},
}

@misc{vanheukelum2026intrinsicalignmentdisksellipticals,
      title={Intrinsic alignment of disks and ellipticals across hydrodynamical simulations}, 
      author={M. L. van Heukelum and N. E. Chisari},
      year={2026},
      eprint={2510.11118},
      archivePrefix={arXiv},
      primaryClass={astro-ph.GA},
      url={https://arxiv.org/abs/2510.11118}, 
}

@article{Hearin_2022,
   title={Differentiable Predictions for Large Scale Structure with SHAMNet},
   volume={5},
   ISSN={2565-6120},
   url={http://dx.doi.org/10.21105/astro.2112.08423},
   DOI={10.21105/astro.2112.08423},
   number={1},
   journal={The Open Journal of Astrophysics},
   publisher={Maynooth University},
   author={Hearin, Andrew P. and Ramachandra, Nesar and Becker, Matthew R. and DeRose, Joseph},
   year={2022},
   month=feb }

@article{Navarro_1996,
   title={The Structure of Cold Dark Matter Halos},
   volume={462},
   ISSN={1538-4357},
   url={http://dx.doi.org/10.1086/177173},
   DOI={10.1086/177173},
   journal={The Astrophysical Journal},
   publisher={American Astronomical Society},
   author={Navarro, Julio F. and Frenk, Carlos S. and White, Simon D. M.},
   year={1996},
   month=may, pages={563} }

@misc{jang2017categoricalreparameterizationgumbelsoftmax,
      title={Categorical Reparameterization with Gumbel-Softmax}, 
      author={Eric Jang and Shixiang Gu and Ben Poole},
      year={2017},
      eprint={1611.01144},
      archivePrefix={arXiv},
      primaryClass={stat.ML},
      url={https://arxiv.org/abs/1611.01144}, 
}

@misc{kingma2022autoencodingvariationalbayes,
      title={Auto-Encoding Variational Bayes}, 
      author={Diederik P Kingma and Max Welling},
      year={2022},
      eprint={1312.6114},
      archivePrefix={arXiv},
      primaryClass={stat.ML},
      url={https://arxiv.org/abs/1312.6114}, 
}

@article{Pandya_2025,
   title={IAEmu: Learning Galaxy Intrinsic Alignment Correlations},
   volume={8},
   ISSN={2565-6120},
   url={http://dx.doi.org/10.33232/001c.151749},
   DOI={10.33232/001c.151749},
   journal={The Open Journal of Astrophysics},
   publisher={Maynooth University},
   author={Pandya, Sneh and Yang, Yuanyuan and Van Alfen, Nicholas and Blazek, Jonathan and Walters, Robin},
   year={2025},
   month=dec }

@article{VanAlfen2025, doi = {10.21105/joss.07421}, url = {https://doi.org/10.21105/joss.07421}, year = {2025}, publisher = {The Open Journal}, volume = {10}, number = {107}, pages = {7421}, author = {Van Alfen, Nicholas and Campbell, Duncan and Hearin, Andrew and Blazek, Jonathan}, title = {Halotools: A New Release Adding Intrinsic Alignments to Halo-Based Methods}, journal = {Journal of Open Source Software}}

@misc{li2022pmwddifferentiablecosmologicalparticlemesh,
      title={pmwd: A Differentiable Cosmological Particle-Mesh $N$-body Library}, 
      author={Yin Li and Libin Lu and Chirag Modi and Drew Jamieson and Yucheng Zhang and Yu Feng and Wenda Zhou and Ngai Pok Kwan and François Lanusse and Leslie Greengard},
      year={2022},
      eprint={2211.09958},
      archivePrefix={arXiv},
      primaryClass={astro-ph.IM},
      url={https://arxiv.org/abs/2211.09958}, 
}

@misc{horowitz2022,
      title={Differentiable Stochastic Halo Occupation Distribution}, 
      author={Benjamin Horowitz and ChangHoon Hahn and Francois Lanusse and Chirag Modi and Simone Ferraro},
      year={2022},
      eprint={2211.03852},
      archivePrefix={arXiv},
      primaryClass={astro-ph.CO},
      url={https://arxiv.org/abs/2211.03852}, 
}

@article{Yuan_2018,
   title={Exploring the squeezed three-point galaxy correlation function with generalized halo occupation distribution models},
   volume={478},
   ISSN={1365-2966},
   url={http://dx.doi.org/10.1093/mnras/sty1089},
   DOI={10.1093/mnras/sty1089},
   number={2},
   journal={Monthly Notices of the Royal Astronomical Society},
   publisher={Oxford University Press (OUP)},
   author={Yuan, Sihan and Eisenstein, Daniel J and Garrison, Lehman H},
   year={2018},
   month=apr, pages={2019–2033} }

@article{Neistein_2011,
   title={A tale of two populations: the stellar mass of central and satellite galaxies: A tale of two populations},
   volume={416},
   ISSN={0035-8711},
   url={http://dx.doi.org/10.1111/j.1365-2966.2011.19145.x},
   DOI={10.1111/j.1365-2966.2011.19145.x},
   number={2},
   journal={Monthly Notices of the Royal Astronomical Society},
   publisher={Oxford University Press (OUP)},
   author={Neistein, Eyal and Li, Cheng and Khochfar, Sadegh and Weinmann, Simone M. and Shankar, Francesco and Boylan-Kolchin, Michael},
   year={2011},
   month=jul, pages={1486–1499} }

@misc{descollaboration2026darkenergysurveyyear,
      title={Dark Energy Survey Year 6 Results: Cosmological Constraints from Galaxy Clustering and Weak Lensing}, 
      author={DES Collaboration and T. M. C. Abbott and M. Adamow and M. Aguena and A. Alarcon and S. S. Allam and O. Alves and A. Amon and D. Anbajagane and F. Andrade-Oliveira and S. Avila and D. Bacon and E. J. Baxter and J. Beas-Gonzalez and K. Bechtol and M. R. Becker and G. M. Bernstein and E. Bertin and J. Blazek and S. Bocquet and D. Brooks and D. Brout and H. Camacho and G. Camacho-Ciurana and R. Camilleri and G. Campailla and A. Campos and A. Carnero Rosell and M. Carrasco Kind and J. Carretero and P. Carrilho and F. J. Castander and R. Cawthon and C. Chang and A. Choi and J. M. Coloma-Nadal and M. Costanzi and M. Crocce and W. d'Assignies and L. N. da Costa and M. E. da Silva Pereira and T. M. Davis and J. De Vicente and J. DeRose and H. T. Diehl and S. Dodelson and P. Doel and C. Doux and A. Drlica-Wagner and T. F. Eifler and J. Elvin-Poole and J. Estrada and S. Everett and A. E. Evrard and J. Fang and A. Farahi and A. Ferté and B. Flaugher and P. Fosalba and J. Frieman and J. García-Bellido and M. Gatti and E. Gaztanaga and G. Giannini and P. Giles and K. Glazebrook and M. Gorsuch and D. Gruen and R. A. Gruendl and J. Gschwend and G. Gutierrez and I. Harrison and W. G. Hartley and E. Henning and K. Herner and S. R. Hinton and D. L. Hollowood and K. Honscheid and E. M. Huff and D. Huterer and B. Jain and D. J. James and M. Jarvis and N. Jeffrey and T. Jeltema and T. Kacprzak and S. Kent and A. Kovacs and E. Krause and R. Kron and K. Kuehn and O. Lahav and S. Lee and E. Legnani and C. Lidman and H. Lin and N. MacCrann and M. Manera and T. Manning and J. L. Marshall and S. Mau and J. McCullough and J. Mena-Fernández and F. Menanteau and R. Miquel and J. J. Mohr and J. Muir and J. Myles and R. C. Nichol and B. Nord and J. H. O'Donnell and R. L. C. Ogando and A. Palmese and M. Paterno and J. Peoples and W. J. Percival and D. Petravick and A. Pieres and A. A. Plazas Malagón and A. Porredon and A. Pourtsidou and J. Prat and C. Preston and M. Raveri and W. Riquelme and M. Rodriguez-Monroy and P. Rogozenski and A. K. Romer and A. Roodman and R. Rosenfeld and A. J. Ross and E. Rozo and E. S. Rykoff and S. Samuroff and C. Sánchez and E. Sanchez and D. Sanchez Cid and T. Schutt and I. Sevilla-Noarbe and E. Sheldon and N. Sherman and T. Shin and M. Smith and M. Soares-Santos and E. Suchyta and M. E. C. Swanson and M. Tabbutt and G. Tarle and D. Thomas and C. To and A. Tong and L. Toribio San Cipriano and M. A. Troxel and M. Tsedrik and D. L. Tucker and V. Vikram and A. R. Walker and N. Weaverdyck and R. H. Wechsler and D. H. Weinberg and J. Weller and V. Wetzell and A. Whyley and R. D. Wilkinson and P. Wiseman and H. -Y. Wu and M. Yamamoto and B. Yanny and B. Yin and G. Zacharegkas and Y. Zhang and J. Zuntz},
      year={2026},
      eprint={2601.14559},
      archivePrefix={arXiv},
      primaryClass={astro-ph.CO},
      url={https://arxiv.org/abs/2601.14559}, 
}

@article{Zheng_2016,
   title={Accurate and efficient halo-based galaxy clustering modelling with simulations},
   volume={458},
   ISSN={1365-2966},
   url={http://dx.doi.org/10.1093/mnras/stw523},
   DOI={10.1093/mnras/stw523},
   number={4},
   journal={Monthly Notices of the Royal Astronomical Society},
   publisher={Oxford University Press (OUP)},
   author={Zheng, Zheng and Guo, Hong},
   year={2016},
   month=mar, pages={4015–4024} }

@article{Hartlap_2006,
   title={Why your model parameter confidences might be too optimistic.  Unbiased estimation of the inverse covariance matrix},
   volume={464},
   ISSN={1432-0746},
   url={http://dx.doi.org/10.1051/0004-6361:20066170},
   DOI={10.1051/0004-6361:20066170},
   number={1},
   journal={Astronomy &amp; Astrophysics},
   publisher={EDP Sciences},
   author={Hartlap, J. and Simon, P. and Schneider, P.},
   year={2006},
   month=dec, pages={399–404} }

@misc{horowitz2025jfofgpuclusterfinding,
      title={jFoF: GPU Cluster Finding with Gradient Propagation}, 
      author={Benjamin Horowitz and Adrian E. Bayer},
      year={2025},
      eprint={2510.26851},
      archivePrefix={arXiv},
      primaryClass={astro-ph.IM},
      url={https://arxiv.org/abs/2510.26851}, 
}

@article{vanalfen_2025_inprep,
    author = {{Van Alfen}, Nicholas and {Blazek}, Jonathan and {Hearin}, Andrew},
    title = "{Coherent 2D and 3D simulations of intrinsic alignments}",
    journal = {in prep.},
    year = {2025}
}

@article{Maion_2024,
   title={HYMALAIA: a hybrid lagrangian model for intrinsic alignments},
   volume={531},
   ISSN={1365-2966},
   url={http://dx.doi.org/10.1093/mnras/stae1331},
   DOI={10.1093/mnras/stae1331},
   number={2},
   journal={Monthly Notices of the Royal Astronomical Society},
   publisher={Oxford University Press (OUP)},
   author={Maion, Francisco and Angulo, Raul E and Bakx, Thomas and Chisari, Nora Elisa and Kurita, Toshiki and Pellejero-Ibáñez, Marcos},
   year={2024},
   month=may, pages={2684–2700} }

@article{Piras_2024,
   title={The future of cosmological likelihood-based inference: accelerated high-dimensional parameter estimation and model comparison},
   volume={7},
   ISSN={2565-6120},
   url={http://dx.doi.org/10.33232/001c.123368},
   DOI={10.33232/001c.123368},
   journal={The Open Journal of Astrophysics},
   publisher={Maynooth University},
   author={Piras, Davide and Polanska, Alicja and Mancini, Alessio Spurio and Price, Matthew A. and McEwen, Jason D.},
   year={2024},
   month=sep }

@misc{dvorkin2022machine,
      title={Machine Learning and Cosmology}, 
      author={Cora Dvorkin and Siddharth Mishra-Sharma and Brian Nord and V. Ashley Villar and Camille Avestruz and Keith Bechtol and Aleksandra Ćiprijanović and Andrew J. Connolly and Lehman H. Garrison and Gautham Narayan and Francisco Villaescusa-Navarro},
      year={2022},
      eprint={2203.08056},
      archivePrefix={arXiv},
      primaryClass={hep-ph}
}

@article{Jagvaral_2022,
   title={Galaxies and haloes on graph neural networks: Deep generative modelling scalar and vector quantities for intrinsic alignment},
   volume={516},
   ISSN={1365-2966},
   url={http://dx.doi.org/10.1093/mnras/stac2083},
   DOI={10.1093/mnras/stac2083},
   number={2},
   journal={Monthly Notices of the Royal Astronomical Society},
   publisher={Oxford University Press (OUP)},
   author={Jagvaral, Yesukhei and Lanusse, François and Singh, Sukhdeep and Mandelbaum, Rachel and Ravanbakhsh, Siamak and Campbell, Duncan},
   year={2022},
   month=aug, pages={2406–2419} }

@misc{nelson2021illustristng,
      title={The IllustrisTNG Simulations: Public Data Release}, 
      author={Dylan Nelson and Volker Springel and Annalisa Pillepich and Vicente Rodriguez-Gomez and Paul Torrey and Shy Genel and Mark Vogelsberger and Ruediger Pakmor and Federico Marinacci and Rainer Weinberger and Luke Kelley and Mark Lovell and Benedikt Diemer and Lars Hernquist},
      year={2021},
      eprint={1812.05609},
      archivePrefix={arXiv},
      primaryClass={astro-ph.GA}
}

@misc{vanalfen_2023,
      title={An Empirical Model For Intrinsic Alignments: Insights From Cosmological Simulations}, 
      author={Nicholas {Van Alfen} and Duncan Campbell and Jonathan Blazek and C. Danielle Leonard and Francois Lanusse and Andrew Hearin and Rachel Mandelbaum and The LSST Dark Energy Science Collaboration},
      year={2024},
      eprint={2311.07374},
      archivePrefix={arXiv},
      primaryClass={astro-ph.CO}
}

@article{Zheng_2007,
	doi = {10.1086/521074},
	url = {https://doi.org/10.1086\%2F521074},
	year = 2007,
	month = {oct},
	publisher = {American Astronomical Society},
	volume = {667},
	number = {2},
	pages = {760--779},
	author = {Zheng Zheng and Alison L. Coil and Idit Zehavi},
	title = {Galaxy Evolution from Halo Occupation Distribution Modeling of {DEEP}2 and {SDSS} Galaxy Clustering},
	journal = {The Astrophysical Journal}
}

@ARTICLE{lsst,
       author = {{Ivezi{\'c}}, {\v{Z}}eljko and {Kahn}, Steven M. and {Tyson}, J. Anthony and {Abel}, Bob and {Acosta}, Emily and {Allsman}, Robyn and {Alonso}, David and {AlSayyad}, Yusra and {Anderson}, Scott F. and {Andrew}, John and {Angel}, James Roger P. and {Angeli}, George Z. and {Ansari}, Reza and {Antilogus}, Pierre and {Araujo}, Constanza and {Armstrong}, Robert and {Arndt}, Kirk T. and {Astier}, Pierre and {Aubourg}, {\'E}ric and {Auza}, Nicole and {Axelrod}, Tim S. and {Bard}, Deborah J. and {Barr}, Jeff D. and {Barrau}, Aurelian and {Bartlett}, James G. and {Bauer}, Amanda E. and {Bauman}, Brian J. and {Baumont}, Sylvain and {Bechtol}, Ellen and {Bechtol}, Keith and {Becker}, Andrew C. and {Becla}, Jacek and {Beldica}, Cristina and {Bellavia}, Steve and {Bianco}, Federica B. and {Biswas}, Rahul and {Blanc}, Guillaume and {Blazek}, Jonathan and {Blandford}, Roger D. and {Bloom}, Josh S. and {Bogart}, Joanne and {Bond}, Tim W. and {Booth}, Michael T. and {Borgland}, Anders W. and {Borne}, Kirk and {Bosch}, James F. and {Boutigny}, Dominique and {Brackett}, Craig A. and {Bradshaw}, Andrew and {Brandt}, William Nielsen and {Brown}, Michael E. and {Bullock}, James S. and {Burchat}, Patricia and {Burke}, David L. and {Cagnoli}, Gianpietro and {Calabrese}, Daniel and {Callahan}, Shawn and {Callen}, Alice L. and {Carlin}, Jeffrey L. and {Carlson}, Erin L. and {Chandrasekharan}, Srinivasan and {Charles-Emerson}, Glenaver and {Chesley}, Steve and {Cheu}, Elliott C. and {Chiang}, Hsin-Fang and {Chiang}, James and {Chirino}, Carol and {Chow}, Derek and {Ciardi}, David R. and {Claver}, Charles F. and {Cohen-Tanugi}, Johann and {Cockrum}, Joseph J. and {Coles}, Rebecca and {Connolly}, Andrew J. and {Cook}, Kem H. and {Cooray}, Asantha and {Covey}, Kevin R. and {Cribbs}, Chris and {Cui}, Wei and {Cutri}, Roc and {Daly}, Philip N. and {Daniel}, Scott F. and {Daruich}, Felipe and {Daubard}, Guillaume and {Daues}, Greg and {Dawson}, William and {Delgado}, Francisco and {Dellapenna}, Alfred and {de Peyster}, Robert and {de Val-Borro}, Miguel and {Digel}, Seth W. and {Doherty}, Peter and {Dubois}, Richard and {Dubois-Felsmann}, Gregory P. and {Durech}, Josef and {Economou}, Frossie and {Eifler}, Tim and {Eracleous}, Michael and {Emmons}, Benjamin L. and {Fausti Neto}, Angelo and {Ferguson}, Henry and {Figueroa}, Enrique and {Fisher-Levine}, Merlin and {Focke}, Warren and {Foss}, Michael D. and {Frank}, James and {Freemon}, Michael D. and {Gangler}, Emmanuel and {Gawiser}, Eric and {Geary}, John C. and {Gee}, Perry and {Geha}, Marla and {Gessner}, Charles J.~B. and {Gibson}, Robert R. and {Gilmore}, D. Kirk and {Glanzman}, Thomas and {Glick}, William and {Goldina}, Tatiana and {Goldstein}, Daniel A. and {Goodenow}, Iain and {Graham}, Melissa L. and {Gressler}, William J. and {Gris}, Philippe and {Guy}, Leanne P. and {Guyonnet}, Augustin and {Haller}, Gunther and {Harris}, Ron and {Hascall}, Patrick A. and {Haupt}, Justine and {Hernandez}, Fabio and {Herrmann}, Sven and {Hileman}, Edward and {Hoblitt}, Joshua and {Hodgson}, John A. and {Hogan}, Craig and {Howard}, James D. and {Huang}, Dajun and {Huffer}, Michael E. and {Ingraham}, Patrick and {Innes}, Walter R. and {Jacoby}, Suzanne H. and {Jain}, Bhuvnesh and {Jammes}, Fabrice and {Jee}, M. James and {Jenness}, Tim and {Jernigan}, Garrett and {Jevremovi{\'c}}, Darko and {Johns}, Kenneth and {Johnson}, Anthony S. and {Johnson}, Margaret W.~G. and {Jones}, R. Lynne and {Juramy-Gilles}, Claire and {Juri{\'c}}, Mario and {Kalirai}, Jason S. and {Kallivayalil}, Nitya J. and {Kalmbach}, Bryce and {Kantor}, Jeffrey P. and {Karst}, Pierre and {Kasliwal}, Mansi M. and {Kelly}, Heather and {Kessler}, Richard and {Kinnison}, Veronica and {Kirkby}, David and {Knox}, Lloyd and {Kotov}, Ivan V. and {Krabbendam}, Victor L. and {Krughoff}, K. Simon and {Kub{\'a}nek}, Petr and {Kuczewski}, John and {Kulkarni}, Shri and {Ku}, John and {Kurita}, Nadine R. and {Lage}, Craig S. and {Lambert}, Ron and {Lange}, Travis and {Langton}, J. Brian and {Le Guillou}, Laurent and {Levine}, Deborah and {Liang}, Ming and {Lim}, Kian-Tat and {Lintott}, Chris J. and {Long}, Kevin E. and {Lopez}, Margaux and {Lotz}, Paul J. and {Lupton}, Robert H. and {Lust}, Nate B. and {MacArthur}, Lauren A. and {Mahabal}, Ashish and {Mandelbaum}, Rachel and {Markiewicz}, Thomas W. and {Marsh}, Darren S. and {Marshall}, Philip J. and {Marshall}, Stuart and {May}, Morgan and {McKercher}, Robert and {McQueen}, Michelle and {Meyers}, Joshua and {Migliore}, Myriam and {Miller}, Michelle and {Mills}, David J. and {Miraval}, Connor and {Moeyens}, Joachim and {Moolekamp}, Fred E. and {Monet}, David G. and {Moniez}, Marc and {Monkewitz}, Serge and {Montgomery}, Christopher and {Morrison}, Christopher B. and {Mueller}, Fritz and {Muller}, Gary P. and {Mu{\~n}oz Arancibia}, Freddy and {Neill}, Douglas R. and {Newbry}, Scott P. and {Nief}, Jean-Yves and {Nomerotski}, Andrei and {Nordby}, Martin and {O'Connor}, Paul and {Oliver}, John and {Olivier}, Scot S. and {Olsen}, Knut and {O'Mullane}, William and {Ortiz}, Sandra and {Osier}, Shawn and {Owen}, Russell E. and {Pain}, Reynald and {Palecek}, Paul E. and {Parejko}, John K. and {Parsons}, James B. and {Pease}, Nathan M. and {Peterson}, J. Matt and {Peterson}, John R. and {Petravick}, Donald L. and {Libby Petrick}, M.~E. and {Petry}, Cathy E. and {Pierfederici}, Francesco and {Pietrowicz}, Stephen and {Pike}, Rob and {Pinto}, Philip A. and {Plante}, Raymond and {Plate}, Stephen and {Plutchak}, Joel P. and {Price}, Paul A. and {Prouza}, Michael and {Radeka}, Veljko and {Rajagopal}, Jayadev and {Rasmussen}, Andrew P. and {Regnault}, Nicolas and {Reil}, Kevin A. and {Reiss}, David J. and {Reuter}, Michael A. and {Ridgway}, Stephen T. and {Riot}, Vincent J. and {Ritz}, Steve and {Robinson}, Sean and {Roby}, William and {Roodman}, Aaron and {Rosing}, Wayne and {Roucelle}, Cecille and {Rumore}, Matthew R. and {Russo}, Stefano and {Saha}, Abhijit and {Sassolas}, Benoit and {Schalk}, Terry L. and {Schellart}, Pim and {Schindler}, Rafe H. and {Schmidt}, Samuel and {Schneider}, Donald P. and {Schneider}, Michael D. and {Schoening}, William and {Schumacher}, German and {Schwamb}, Megan E. and {Sebag}, Jacques and {Selvy}, Brian and {Sembroski}, Glenn H. and {Seppala}, Lynn G. and {Serio}, Andrew and {Serrano}, Eduardo and {Shaw}, Richard A. and {Shipsey}, Ian and {Sick}, Jonathan and {Silvestri}, Nicole and {Slater}, Colin T. and {Smith}, J. Allyn and {Smith}, R. Chris and {Sobhani}, Shahram and {Soldahl}, Christine and {Storrie-Lombardi}, Lisa and {Stover}, Edward and {Strauss}, Michael A. and {Street}, Rachel A. and {Stubbs}, Christopher W. and {Sullivan}, Ian S. and {Sweeney}, Donald and {Swinbank}, John D. and {Szalay}, Alexander and {Takacs}, Peter and {Tether}, Stephen A. and {Thaler}, Jon J. and {Thayer}, John Gregg and {Thomas}, Sandrine and {Thornton}, Adam J. and {Thukral}, Vaikunth and {Tice}, Jeffrey and {Trilling}, David E. and {Turri}, Max and {Van Berg}, Richard and {Vanden Berk}, Daniel and {Vetter}, Kurt and {Virieux}, Francoise and {Vucina}, Tomislav and {Wahl}, William and {Walkowicz}, Lucianne and {Walsh}, Brian and {Walter}, Christopher W. and {Wang}, Daniel L. and {Wang}, Shin-Yawn and {Warner}, Michael and {Wiecha}, Oliver and {Willman}, Beth and {Winters}, Scott E. and {Wittman}, David and {Wolff}, Sidney C. and {Wood-Vasey}, W. Michael and {Wu}, Xiuqin and {Xin}, Bo and {Yoachim}, Peter and {Zhan}, Hu},
        title = "{LSST: From Science Drivers to Reference Design and Anticipated Data Products}",
      journal = {\apj},
         year = 2019,
        month = mar,
       volume = {873},
       number = {2},
          eid = {111},
        pages = {111},
          doi = {10.3847/1538-4357/ab042c},
archivePrefix = {arXiv},
       eprint = {0805.2366},
 primaryClass = {astro-ph},
       adsurl = {https://ui.adsabs.harvard.edu/abs/2019ApJ...873..111I}
}

@article{euclid,
   title={Euclid preparation: I. The Euclid Wide Survey},
   volume={662},
   ISSN={1432-0746},
   url={http://dx.doi.org/10.1051/0004-6361/202141938},
   DOI={10.1051/0004-6361/202141938},
   journal={Astronomy &amp; Astrophysics},
   publisher={EDP Sciences},
   author={Scaramella, R. and Amiaux, J. and Mellier, Y. and Burigana, C. and Carvalho, C. S. and Cuillandre, J.-C. and Da Silva, A. and Derosa, A. and Dinis, J. and Maiorano, E. and Maris, M. and Tereno, I. and Laureijs, R. and Boenke, T. and Buenadicha, G. and Dupac, X. and Gaspar Venancio, L. M. and Gómez-Álvarez, P. and Hoar, J. and Lorenzo Alvarez, J. and Racca, G. D. and Saavedra-Criado, G. and Schwartz, J. and Vavrek, R. and Schirmer, M. and Aussel, H. and Azzollini, R. and Cardone, V. F. and Cropper, M. and Ealet, A. and Garilli, B. and Gillard, W. and Granett, B. R. and Guzzo, L. and Hoekstra, H. and Jahnke, K. and Kitching, T. and Maciaszek, T. and Meneghetti, M. and Miller, L. and Nakajima, R. and Niemi, S. M. and Pasian, F. and Percival, W. J. and Pottinger, S. and Sauvage, M. and Scodeggio, M. and Wachter, S. and Zacchei, A. and Aghanim, N. and Amara, A. and Auphan, T. and Auricchio, N. and Awan, S. and Balestra, A. and Bender, R. and Bodendorf, C. and Bonino, D. and Branchini, E. and Brau-Nogue, S. and Brescia, M. and Candini, G. P. and Capobianco, V. and Carbone, C. and Carlberg, R. G. and Carretero, J. and Casas, R. and Castander, F. J. and Castellano, M. and Cavuoti, S. and Cimatti, A. and Cledassou, R. and Congedo, G. and Conselice, C. J. and Conversi, L. and Copin, Y. and Corcione, L. and Costille, A. and Courbin, F. and Degaudenzi, H. and Douspis, M. and Dubath, F. and Duncan, C. A. J. and Dusini, S. and Farrens, S. and Ferriol, S. and Fosalba, P. and Fourmanoit, N. and Frailis, M. and Franceschi, E. and Franzetti, P. and Fumana, M. and Gillis, B. and Giocoli, C. and Grazian, A. and Grupp, F. and Haugan, S. V. H. and Holmes, W. and Hormuth, F. and Hudelot, P. and Kermiche, S. and Kiessling, A. and Kilbinger, M. and Kohley, R. and Kubik, B. and Kümmel, M. and Kunz, M. and Kurki-Suonio, H. and Lahav, O. and Ligori, S. and Lilje, P. B. and Lloro, I. and Mansutti, O. and Marggraf, O. and Markovic, K. and Marulli, F. and Massey, R. and Maurogordato, S. and Melchior, M. and Merlin, E. and Meylan, G. and Mohr, J. J. and Moresco, M. and Morin, B. and Moscardini, L. and Munari, E. and Nichol, R. C. and Padilla, C. and Paltani, S. and Peacock, J. and Pedersen, K. and Pettorino, V. and Pires, S. and Poncet, M. and Popa, L. and Pozzetti, L. and Raison, F. and Rebolo, R. and Rhodes, J. and Rix, H.-W. and Roncarelli, M. and Rossetti, E. and Saglia, R. and Schneider, P. and Schrabback, T. and Secroun, A. and Seidel, G. and Serrano, S. and Sirignano, C. and Sirri, G. and Skottfelt, J. and Stanco, L. and Starck, J. L. and Tallada-Crespí, P. and Tavagnacco, D. and Taylor, A. N. and Teplitz, H. I. and Toledo-Moreo, R. and Torradeflot, F. and Trifoglio, M. and Valentijn, E. A. and Valenziano, L. and Verdoes Kleijn, G. A. and Wang, Y. and Welikala, N. and Weller, J. and Wetzstein, M. and Zamorani, G. and Zoubian, J. and Andreon, S. and Baldi, M. and Bardelli, S. and Boucaud, A. and Camera, S. and Di Ferdinando, D. and Fabbian, G. and Farinelli, R. and Galeotta, S. and Graciá-Carpio, J. and Maino, D. and Medinaceli, E. and Mei, S. and Neissner, C. and Polenta, G. and Renzi, A. and Romelli, E. and Rosset, C. and Sureau, F. and Tenti, M. and Vassallo, T. and Zucca, E. and Baccigalupi, C. and Balaguera-Antolínez, A. and Battaglia, P. and Biviano, A. and Borgani, S. and Bozzo, E. and Cabanac, R. and Cappi, A. and Casas, S. and Castignani, G. and Colodro-Conde, C. and Coupon, J. and Courtois, H. M. and Cuby, J. and de la Torre, S. and Desai, S. and Dole, H. and Fabricius, M. and Farina, M. and Ferreira, P. G. and Finelli, F. and Flose-Reimberg, P. and Fotopoulou, S. and Ganga, K. and Gozaliasl, G. and Hook, I. M. and Keihanen, E. and Kirkpatrick, C. C. and Liebing, P. and Lindholm, V. and Mainetti, G. and Martinelli, M. and Martinet, N. and Maturi, M. and McCracken, H. J. and Metcalf, R. B. and Morgante, G. and Nightingale, J. and Nucita, A. and Patrizii, L. and Potter, D. and Riccio, G. and Sánchez, A. G. and Sapone, D. and Schewtschenko, J. A. and Schultheis, M. and Scottez, V. and Teyssier, R. and Tutusaus, I. and Valiviita, J. and Viel, M. and Vriend, W. and Whittaker, L.},
   year={2022},
   month=jun, pages={A112} }

@misc{berman2025softclustering,
      title={On Soft Clustering For Correlation Estimators: Model Uncertainty, Differentiability, and Surrogates}, 
      author={Edward Berman and Sneh Pandya and Jacqueline McCleary and Marko Shuntov and Caitlin Casey and Nicole Drakos and Andreas Faisst and Steven Gillman and Ghassem Gozaliasl and Natalie Hogg and Jeyhan Kartaltepe and Anton Koekemoer and Wilfried Mercier and Diana Scognamiglio and COSMOS-Web and : and The JWST Cosmic Origins Survey},
      year={2025},
      eprint={2504.06174},
      archivePrefix={arXiv},
      primaryClass={astro-ph.IM},
      url={https://arxiv.org/abs/2504.06174}, 
}

@misc{roman,
      title={The Wide Field Infrared Survey Telescope: 100 Hubbles for the 2020s}, 
      author={Rachel Akeson and Lee Armus and Etienne Bachelet and Vanessa Bailey and Lisa Bartusek and Andrea Bellini and Dominic Benford and David Bennett and Aparna Bhattacharya and Ralph Bohlin and Martha Boyer and Valerio Bozza and Geoffrey Bryden and Sebastiano Calchi Novati and Kenneth Carpenter and Stefano Casertano and Ami Choi and David Content and Pratika Dayal and Alan Dressler and Olivier Doré and S. Michael Fall and Xiaohui Fan and Xiao Fang and Alexei Filippenko and Steven Finkelstein and Ryan Foley and Steven Furlanetto and Jason Kalirai and B. Scott Gaudi and Karoline Gilbert and Julien Girard and Kevin Grady and Jenny Greene and Puragra Guhathakurta and Chen Heinrich and Shoubaneh Hemmati and David Hendel and Calen Henderson and Thomas Henning and Christopher Hirata and Shirley Ho and Eric Huff and Anne Hutter and Rolf Jansen and Saurabh Jha and Samson Johnson and David Jones and Jeremy Kasdin and Patrick Kelly and Robert Kirshner and Anton Koekemoer and Jeffrey Kruk and Nikole Lewis and Bruce Macintosh and Piero Madau and Sangeeta Malhotra and Kaisey Mandel and Elena Massara and Daniel Masters and Julie McEnery and Kristen McQuinn and Peter Melchior and Mark Melton and Bertrand Mennesson and Molly Peeples and Matthew Penny and Saul Perlmutter and Alice Pisani and Andrés Plazas and Radek Poleski and Marc Postman and Clément Ranc and Bernard Rauscher and Armin Rest and Aki Roberge and Brant Robertson and Steven Rodney and James Rhoads and Jason Rhodes and Russell Ryan Jr. au2 and Kailash Sahu and David Sand and Dan Scolnic and Anil Seth and Yossi Shvartzvald and Karelle Siellez and Arfon Smith and David Spergel and Keivan Stassun and Rachel Street and Louis-Gregory Strolger and Alexander Szalay and John Trauger and M. A. Troxel and Margaret Turnbull and Roeland van der Marel and Anja von der Linden and Yun Wang and David Weinberg and Benjamin Williams and Rogier Windhorst and Edward Wollack and Hao-Yi Wu and Jennifer Yee and Neil Zimmerman},
      year={2019},
      eprint={1902.05569},
      archivePrefix={arXiv},
      primaryClass={astro-ph.IM},
      url={https://arxiv.org/abs/1902.05569}, 
}

@misc{hoffman2011nouturnsampleradaptivelysetting,
      title={The No-U-Turn Sampler: Adaptively Setting Path Lengths in Hamiltonian Monte Carlo}, 
      author={Matthew D. Hoffman and Andrew Gelman},
      year={2011},
      eprint={1111.4246},
      archivePrefix={arXiv},
      primaryClass={stat.CO},
      url={https://arxiv.org/abs/1111.4246}, 
}

@misc{
jagvaral2023diffusion,
title={{DIFFUSION} {GENERATIVE} {MODELS} {ON} {SO}(3)},
author={Yesukhei Jagvaral and Francois Lanusse and Rachel Mandelbaum},
year={2023},
url={https://openreview.net/forum?id=jHA-yCyBGb}
}

@article{Hirata_2004,
	Adsurl = {http://adsabs.harvard.edu/abs/2004PhRvD..70f3526H},
	Author = {{Hirata}, C.~M. and {Seljak}, U.},
	Doi = {10.1103/PhysRevD.70.063526},
	Eprint = {astro-ph/0406275},
	Journal = {\prd},
	Month = sep,
	Number = 6,
	Pages = {063526-+},
	Title = {{Intrinsic alignment-lensing interference as a contaminant of cosmic shear}},
	Volume = 70,
	Year = 2004}

@article{Bridle_2007,
	Adsurl = {http://adsabs.harvard.edu/abs/2007NJPh....9..444B},
	Archiveprefix = {arXiv},
	Author = {{Bridle}, S. and {King}, L.},
	Doi = {10.1088/1367-2630/9/12/444},
	Eprint = {0705.0166},
	Journal = {New Journal of Physics},
	Month = dec,
	Pages = {444},
	Title = {{Dark energy constraints from cosmic shear power spectra: impact of intrinsic alignments on photometric redshift requirements}},
	Volume = 9,
	Year = 2007}

@article{Blazek_2019,
  title = {Beyond linear galaxy alignments},
  author = {Blazek, Jonathan A. and MacCrann, Niall and Troxel, M. A. and Fang, Xiao},
  journal = {Phys. Rev. D},
  volume = {100},
  issue = {10},
  pages = {103506},
  numpages = {19},
  year = {2019},
  month = {Nov},
  publisher = {American Physical Society},
  doi = {10.1103/PhysRevD.100.103506},
  url = {https://link.aps.org/doi/10.1103/PhysRevD.100.103506}
}

@ARTICLE{vlah_2020,
       author = {{Vlah}, Zvonimir and {Chisari}, Nora Elisa and {Schmidt}, Fabian},
        title = "{An EFT description of galaxy intrinsic alignments}",
      journal = {\jcap},
         year = 2020,
        month = jan,
       volume = {2020},
       number = {1},
          eid = {025},
        pages = {025},
          doi = {10.1088/1475-7516/2020/01/025},
archivePrefix = {arXiv},
       eprint = {1910.08085},
 primaryClass = {astro-ph.CO},
       adsurl = {https://ui.adsabs.harvard.edu/abs/2020JCAP...01..025V}
}

@ARTICLE{Hoffmann_2022,
       author = {{Hoffmann}, Kai and {Secco}, Lucas F. and {Blazek}, Jonathan and {Crocce}, Martin and {Tallada-Cresp{\'\i}}, Pau and {Samuroff}, Simon and {Prat}, Judit and {Carretero}, Jorge and {Fosalba}, Pablo and {Gazta{\~n}aga}, Enrique and {Castander}, Francisco J. and {DES Collaboration}},
        title = "{Modeling intrinsic galaxy alignment in the MICE simulation}",
      journal = {\prd},
         year = 2022,
        month = dec,
       volume = {106},
       number = {12},
          eid = {123510},
        pages = {123510},
          doi = {10.1103/PhysRevD.106.123510},
archivePrefix = {arXiv},
       eprint = {2206.14219},
 primaryClass = {astro-ph.CO},
       adsurl = {https://ui.adsabs.harvard.edu/abs/2022PhRvD.106l3510H}
}

@article{Blazek_2015,
doi = {10.1088/1475-7516/2015/08/015},
url = {https://dx.doi.org/10.1088/1475-7516/2015/08/015},
year = {2015},
month = {aug},
publisher = {},
volume = {2015},
number = {08},
pages = {015},
author = {Jonathan Blazek and Zvonimir Vlah and Uroš Seljak},
title = {Tidal alignment of galaxies},
journal = {Journal of Cosmology and Astroparticle Physics}
}

@ARTICLE{vlah_2021,
       author = {{Vlah}, Zvonimir and {Chisari}, Nora Elisa and {Schmidt}, Fabian},
        title = "{Galaxy shape statistics in the effective field theory}",
      journal = {\jcap},
         year = 2021,
        month = may,
       volume = {2021},
       number = {5},
          eid = {061},
        pages = {061},
          doi = {10.1088/1475-7516/2021/05/061},
archivePrefix = {arXiv},
       eprint = {2012.04114},
 primaryClass = {astro-ph.CO},
       adsurl = {https://ui.adsabs.harvard.edu/abs/2021JCAP...05..061V}
}

@ARTICLE{jagvaral24_arxiv,
       author = {{Jagvaral}, Yesukhei and {Lanusse}, Francois and {Mandelbaum}, Rachel},
        title = "{Geometric deep learning for galaxy-halo connection: a case study for galaxy intrinsic alignments}",
      journal = {arXiv e-prints},
         year = 2024,
        month = sep,
          eid = {arXiv:2409.18761},
        pages = {arXiv:2409.18761},
          doi = {10.48550/arXiv.2409.18761},
archivePrefix = {arXiv},
       eprint = {2409.18761},
 primaryClass = {astro-ph.GA},
       adsurl = {https://ui.adsabs.harvard.edu/abs/2024arXiv240918761J}
}

@article{tenneti16,
	Adsurl = {http://adsabs.harvard.edu/abs/2016MNRAS.462.2668T},
	Archiveprefix = {arXiv},
	Author = {{Tenneti}, A. and {Mandelbaum}, R. and {Di Matteo}, T.},
	Doi = {10.1093/mnras/stw1823},
	Eprint = {1510.07024},
	Journal = {\mnras},
	Month = nov,
	Pages = {2668-2680},
	Title = {{Intrinsic alignments of disc and elliptical galaxies in the MassiveBlack-II and Illustris simulations}},
	Volume = 462,
	Year = 2016}

@ARTICLE{samuroff21,
       author = {{Samuroff}, S. and {Mandelbaum}, R. and {Blazek}, J.},
        title = "{Advances in constraining intrinsic alignment models with hydrodynamic simulations}",
      journal = {\mnras},
         year = 2021,
        month = nov,
       volume = {508},
       number = {1},
        pages = {637-664},
          doi = {10.1093/mnras/stab2520},
archivePrefix = {arXiv},
       eprint = {2009.10735},
       adsurl = {https://ui.adsabs.harvard.edu/abs/2021MNRAS.508..637S}
}

@ARTICLE{fortuna21a,
       author = {{Fortuna}, Maria Cristina and {Hoekstra}, Henk and {Joachimi}, Benjamin and {Johnston}, Harry and {Chisari}, Nora Elisa and {Georgiou}, Christos and {Mahony}, Constance},
        title = "{The halo model as a versatile tool to predict intrinsic alignments}",
      journal = {\mnras},
         year = 2021,
        month = feb,
       volume = {501},
       number = {2},
        pages = {2983-3002},
          doi = {10.1093/mnras/staa3802},
archivePrefix = {arXiv},
       eprint = {2003.02700},
 primaryClass = {astro-ph.CO},
       adsurl = {https://ui.adsabs.harvard.edu/abs/2021MNRAS.501.2983F}
}

@article{Peebles1980,
  author = {Peebles, P. J. E.},
  title = {The Large-Scale Structure of the Universe},
  journal = {Princeton University Press},
  year = {1980}
}

@article{Davis1983,
  author = {Davis, Marc and Peebles, P. J. E.},
  title = {A survey of galaxy redshifts. V. The two-point position and velocity correlations},
  journal = {The Astrophysical Journal},
  volume = {267},
  pages = {465--482},
  year = {1983}
}

@article{Troxel2015,
  author = {Troxel, M. A. and Ishak, Mustapha},
  title = {The intrinsic alignment of galaxies and its impact on weak gravitational lensing in an era of precision cosmology},
  journal = {Physics Reports},
  volume = {558},
  pages = {1--59},
  year = {2015}
}

@article{Joachimi2015,
  author = {Joachimi, B. and others},
  title = {Galaxy alignments: An overview},
  journal = {Space Science Reviews},
  volume = {193},
  pages = {1--65},
  year = {2015}
}

@article{Kiessling2015,
  author = {Kiessling, A. and others},
  title = {Galaxy alignments: Theory, modelling and simulations},
  journal = {Space Science Reviews},
  volume = {193},
  pages = {67--136},
  year = {2015}
}

@article{Peacock2000,
  author = {Peacock, J. A. and Smith, R. E.},
  title = {Halo occupation numbers and galaxy bias},
  journal = {Monthly Notices of the Royal Astronomical Society},
  volume = {318},
  pages = {1144--1156},
  year = {2000}
}

@article{Seljak2000,
  author = {Seljak, Uro{\v{s}}},
  title = {Analytic model for galaxy and dark matter clustering},
  journal = {Monthly Notices of the Royal Astronomical Society},
  volume = {318},
  pages = {203--213},
  year = {2000}
}

@article{Berlind2002,
  author = {Berlind, Andreas A. and Weinberg, David H.},
  title = {The Halo Occupation Distribution: Toward an Empirical Determination of the Relation between Galaxies and Mass},
  journal = {The Astrophysical Journal},
  volume = {575},
  pages = {587--616},
  year = {2002}
}

@article{Zheng2005,
  author = {Zheng, Zheng and others},
  title = {Theoretical Models of the Halo Occupation Distribution: Separating Central and Satellite Galaxies},
  journal = {The Astrophysical Journal},
  volume = {633},
  pages = {791--809},
  year = {2005}
}

@article{Zehavi2011,
  author = {Zehavi, Idit and others},
  title = {Galaxy Clustering in the Completed SDSS Redshift Survey: The Dependence on Color and Luminosity},
  journal = {The Astrophysical Journal},
  volume = {736},
  pages = {59},
  year = {2011}
}

@article{Coupon2012,
  author = {Coupon, J. and others},
  title = {Galaxy clustering in the CFHTLS-Wide: the changing relationship between galaxies and haloes since z ~ 1.2},
  journal = {Astronomy \& Astrophysics},
  volume = {542},
  pages = {A5},
  year = {2012}
}

@article{Parejko2013,
  author = {Parejko, John K. and others},
  title = {The clustering of galaxies in the SDSS-III Baryon Oscillation Spectroscopic Survey: the low-redshift sample},
  journal = {Monthly Notices of the Royal Astronomical Society},
  volume = {429},
  pages = {98--112},
  year = {2013}
}

@article{Zheng2009,
  author = {Zheng, Zheng and Coil, Alison L. and Zehavi, Idit},
  title = {Galaxy Evolution from Halo Occupation Distribution Modeling of DEEP2 and SDSS Galaxy Clustering},
  journal = {The Astrophysical Journal},
  volume = {667},
  pages = {760--779},
  year = {2007}
}

@article{Hearin2016,
  author = {Hearin, Andrew P. and others},
  title = {Introducing decorated HODs: modelling assembly bias in the galaxy-halo connection},
  journal = {Monthly Notices of the Royal Astronomical Society},
  volume = {460},
  pages = {2552--2570},
  year = {2016}
}

@article{Wechsler2018,
  author = {Wechsler, Risa H. and Tinker, Jeremy L.},
  title = {The Connection Between Galaxies and Their Dark Matter Halos},
  journal = {Annual Review of Astronomy and Astrophysics},
  volume = {56},
  pages = {435--487},
  year = {2018}
}

@article{Schneider2010,
  author = {Schneider, Michael D. and Bridle, Sarah},
  title = {A halo model for intrinsic alignments of galaxy ellipticities},
  journal = {Monthly Notices of the Royal Astronomical Society},
  volume = {402},
  pages = {2127--2139},
  year = {2010}
}

@article{Joachimi2013a,
  author = {Joachimi, B. and others},
  title = {Galaxy alignments: An overview},
  journal = {Monthly Notices of the Royal Astronomical Society},
  volume = {431},
  pages = {477--492},
  year = {2013}
}

@article{Joachimi2013b,
  author = {Joachimi, B. and others},
  title = {Constraints on intrinsic alignment contamination of weak lensing surveys using the MegaZ-LRG sample},
  journal = {Monthly Notices of the Royal Astronomical Society},
  volume = {436},
  pages = {819--838},
  year = {2013}
}

@article{Hearin2017,
  author = {Hearin, Andrew P. and others},
  title = {Forward Modeling of Large-scale Structure: An Open-source Approach with Halotools},
  journal = {The Astronomical Journal},
  volume = {154},
  pages = {190},
  year = {2017}
}

@article{Duane1987,
  author = {Duane, Simon and Kennedy, A. D. and Pendleton, Brian J. and Roweth, Duncan},
  title = {Hybrid Monte Carlo},
  journal = {Physics Letters B},
  volume = {195},
  pages = {216--222},
  year = {1987}
}

@article{Neal2011,
  author = {Neal, Radford M.},
  title = {MCMC using Hamiltonian dynamics},
  journal = {Handbook of Markov Chain Monte Carlo},
  volume = {2},
  pages = {113--162},
  year = {2011}
}

@article{Maddison2017,
  author = {Maddison, Chris J. and Mnih, Andriy and Teh, Yee Whye},
  title = {The Concrete Distribution: A Continuous Relaxation of Discrete Random Variables},
  journal = {arXiv e-prints},
  eprint = {1611.00712},
  year = {2017}
}

@article{Klypin2016,
  author = {Klypin, Anatoly and others},
  title = {MultiDark simulations: the story of dark matter halo concentrations and density profiles},
  journal = {Monthly Notices of the Royal Astronomical Society},
  volume = {457},
  pages = {4340--4359},
  year = {2016}
}

@article{phan2019composable,
  title={Composable Effects for Flexible and Accelerated Probabilistic Programming in NumPyro},
  author={Phan, Du and Pradhan, Neeraj and Jankowiak, Martin},
  journal={arXiv preprint arXiv:1912.11554},
  year={2019}
}

@article{Rodriguez-Puebla2016,
  author = {Rodr{\'\i}guez-Puebla, Aldo and others},
  title = {The stellar-to-halo mass relation of local galaxies segregates by color},
  journal = {Monthly Notices of the Royal Astronomical Society},
  volume = {462},
  pages = {893--916},
  year = {2016}
}

@misc{kingma2017adammethodstochasticoptimization,
      title={Adam: A Method for Stochastic Optimization}, 
      author={Diederik P. Kingma and Jimmy Ba},
      year={2017},
      eprint={1412.6980},
      archivePrefix={arXiv},
      primaryClass={cs.LG},
      url={https://arxiv.org/abs/1412.6980}, 
}

@software{jax2018github,
  author = {James Bradbury and Roy Frostig and Peter Hawkins and Matthew James Johnson and Chris Leary and Dougal Maclaurin and George Necula and Adam Paszke and Jake Vander{P}las and Skye Wanderman-{M}ilne and Qiao Zhang},
  title = {{JAX}: composable transformations of {P}ython+{N}um{P}y programs},
  url = {http://github.com/jax-ml/jax},
  version = {0.3.13},
  year = {2018},
}

@Article{Watson:1965:EDS,
  author =       "G. S. Watson",
  title =        "Equatorial Distributions on a Sphere",
  journal =      "j-BIOMETRIKA",
  volume =       "52",
  number =       "1/2",
  pages =        "193--201",
  month =        jun,
  year =         "1965",
  CODEN =        "BIOKAX",
  DOI =          "https://doi.org/10.2307/2333824",
  ISSN =         "0006-3444 (print), 1464-3510 (electronic)",
  ISSN-L =       "0006-3444",
  MRclass =      "62.20 (62.25)",
  MRnumber =     "0207115 (34 \#6931)",
  MRreviewer =   "P. A. P. Moran",
  bibdate =      "Sat Jun 21 14:33:18 MDT 2014",
  bibsource =    "http://www.jstor.org/journals/00063444.html;
                 http://www.jstor.org/stable/i315451;
                 https://www.math.utah.edu/pub/tex/bib/biometrika1960.bib",
  URL =          "http://www.jstor.org/stable/2333824",
  acknowledgement = ack-nhfb,
  fjournal =     "Biometrika",
  journal-URL =  "http://biomet.oxfordjournals.org/content/by/year;
                 http://www.jstor.org/journals/00063444.html",
}

\FloatBarrier
\newpage\null\newpage
\newpage\null\newpage
\appendix

\section{Computational Efficiency}
\label{app:efficiency}

\begin{figure}
    \centering
    \includegraphics[width=\columnwidth]{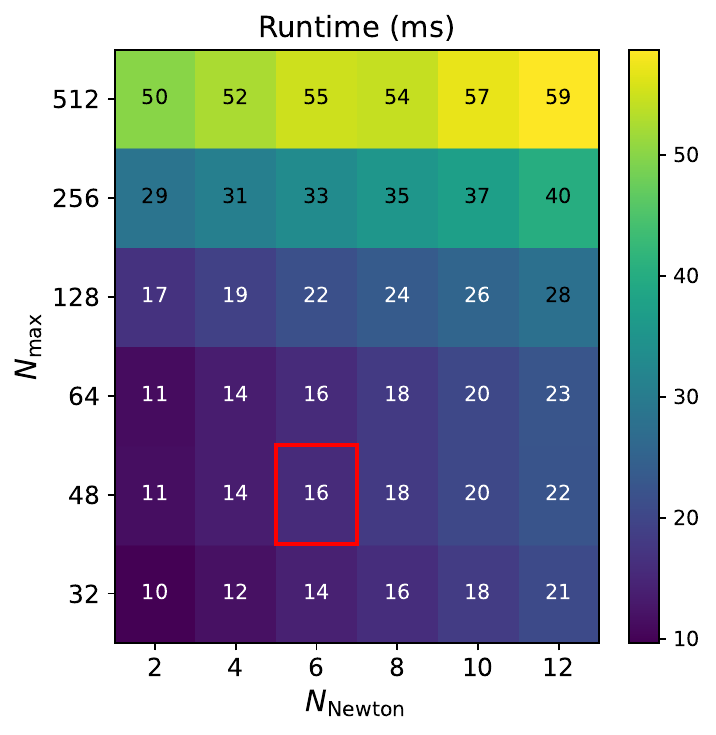}
    % \caption{Benchmarking the computational performance and approximation accuracy of the differentiable satellite occupation and Dimroth--Watson sampling procedures. \textbf{Left panel:} Runtime in milliseconds for the combined satellite sampling and orientation assignment pipeline as a function of $N_{\rm max}$ (the number of Bernoulli trials used in the binomial approximation to Poisson satellite sampling) and $N_{\rm Newton}$ (the number of Newton iterations for inverse-CDF sampling from the Dimroth--Watson distribution). The red box indicates the fiducial configuration used throughout this work ($N_{\rm max} = 48$, $N_{\rm Newton} = 6$). Runtime scales approximately linearly with $N_{\rm max}$, while dependence on $N_{\rm Newton}$ is weaker. \textbf{Center panel:} Total variation distance between the binomial approximation and the target Poisson distribution, averaged over $\lambda \in [0.5, 10]$. The approximation error decreases with increasing $N_{\rm max}$, with the fiducial $N_{\rm max} = 48$ achieving a TV distance of $\sim 0.03$. \textbf{Right panel:} Mean absolute error in the sampled $\cos\theta$ values from the Newton solver relative to a high-precision reference ($N_{\rm Newton} = 50$), shown on a logarithmic scale. Convergence is rapid: $N_{\rm Newton} = 6$ achieves errors of $\sim 10^{-7}$, while $N_{\rm Newton} \geq 10$ reaches machine precision. These benchmarks use $10{,}000$ synthetic host halos and $50{,}000$ satellite orientation samples, averaged over 10 runs.}
    \caption{Runtime in milliseconds for the combined satellite sampling and orientation assignment pipeline as a function of $N_{\rm max}$ and $N_{\rm Newton}$. The red box indicates the fiducial configuration for both parameters used throughout this work. Runtime scales approximately linearly with $N_{\rm max}$, while dependence on $N_{\rm Newton}$ is weaker. Benchmarks use $10{,}000$ synthetic host halos and $50{,}000$ satellite orientation samples, averaged over 50 runs.}
    \label{fig:efficiency}
\end{figure}

We benchmark the computational performance of \diffhodia\ by comparing the relative costs associated with $N_{\rm max}$, used in the differentiable satellite occupation, and $N_{\rm Newton}$, used for inverse CDF sampling from the Dimroth–-Watson distribution. This analysis focuses on the isolated cost of these computations, not the full \diffhodia\ pipeline. We use a synthetic dataset consisting of 10{,}000 host halos and 50{,}000 satellite orientation samples, averaged over 50 runs, executed on a 2021 M1 Max MacBook Pro. The results are shown in Figure~\ref{fig:efficiency}.

We find that satellite sampling dominates the computational cost, with runtime scaling approximately linearly with $N_{\rm max}$, as illustrated by the color gradient in Figure~\ref{fig:efficiency}.
Although increasing $N_{\rm max}$ improves the validity of the Binomial–Poisson approximation (Equation~\ref{eq:binomial_variance}), our fiducial choice of $N_{\rm max} = 48$ yields excellent agreement with \halotoolsia, as quantified by the galaxy clustering correlation function in Figure~\ref{fig:fullcomparison}.
As similarly noted by \citet{horowitz2022}, increasing $N_{\rm max}$ beyond this value does not improve the accuracy of relevant summary statistics.

We additionally investigate how varying $N_{\rm Newton}$ impacts the accuracy of the differentiable Dimroth--Watson procedure.
Using the fiducial alignment parameters introduced earlier, we compute the percent error in the empirical expectation value of the misalignment angle, $\langle \cos^2 \theta \rangle$, relative to a high-precision reference computed with $N_{\rm Newton} = 50$.
We find that the bias decreases with increasing $N_{\rm Newton}$: at $N_{\rm Newton} = 2$, the bias is 4.91\%, dropping to 0.0365\% at $N_{\rm Newton} = 4$, and further to $3.4 \times 10^{-9}$\% at $N_{\rm Newton} = 6$.
For $N_{\rm Newton} \geq 8$, the bias becomes numerically indistinguishable from zero.

\section{Differentiable Galaxy Clustering Statistics}
\label{app:correlations}

\begin{figure*}
    \centering
    \includegraphics[width=.9\textwidth]{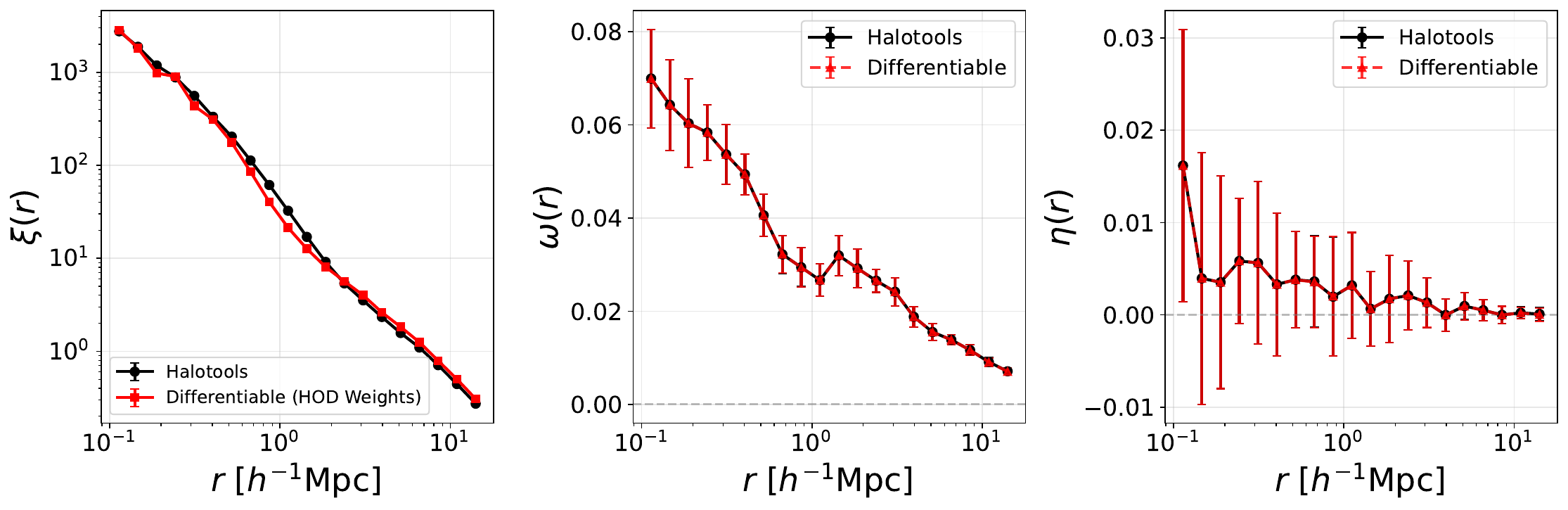}
    \caption{Validation of differentiable correlation function estimators against \texttt{halotools} reference implementations, averaged over 20 independent catalog realizations with error bars indicating the standard deviation across realizations. \textbf{Left:} Galaxy position-position correlation function $\xi(r)$. The black circles show the (non-differentiable) measurement from \texttt{halotools.mock\_observables.tpcf}, while red squares show the differentiable estimator using HOD-derived occupation probability weights. \textbf{Center:} Galaxy position-orientation correlation function $\omega(r)$, comparing \texttt{halotools.mock\_observables.ed\_3d} (black) with our differentiable estimator (red). \textbf{Right:} Galaxy orientation-orientation correlation function $\eta(r)$, comparing \texttt{halotools.mock\_observables.ee\_3d} (black) with our differentiable estimator (red). Both $\omega(r)$ and $\eta(r)$ show excellent agreement between implementations, with $\eta(r)$ exhibiting larger statistical fluctuations due to galaxy shape noise. 
    The differentiable estimators enable gradient-based inference: $\xi(r)$ gradients flow through galaxy occupation weights from HOD parameters, while $\omega(r)$ and $\eta(r)$ gradients flow through orientation vectors from IA parameters.
    }
    \label{fig:diffcor}
\end{figure*}

2PCF calculations require galaxy pair counting, which is a discrete operation and not easily made differentiable.
To compute 2PCFs differentiably, we follow the methodology of \cite{Hearin_2022}.
This requires treating galaxies as existing with a probability $p$, as opposed to discrete objects with weight 1 (exists) or 0 (does not exist).

For central galaxies, the weight $w_i^{\rm cen}$ corresponds to the mean central occupation probability from Equation~\ref{eqn:ncen}:
\begin{equation}
w_i^{\rm cen} = \langle N_{\rm cen}(M) \rangle,
\end{equation}
evaluated at the host halo mass $M$.
For satellite galaxies placed in subhalos, the weight is computed as follows:
\begin{equation}
w_i^{\rm sat} = q_i \cdot \langle N_{\rm sat}(M) \rangle,
\end{equation}
where $q_i$ is the softmax weight prioritizing massive subhalos, defined in Equation \ref{eqn:softmaxweight}.
These weights encode the probability that each galaxy exists in the catalog, enabling gradient flow through the HOD parameters to clustering statistics like $\xi(r)$. Orientation-dependent statistics $\omega(r)$ and $\eta(r)$ receive gradients through the IA parameters.

$\xi(r)$ quantifies the excess probability of finding galaxy pairs at separation $r$ relative to a uniform random distribution:
\begin{equation}
\xi(r) = \frac{DD(r)}{RR(r)} - 1,
\end{equation}
where $DD(r)$ is the (weighted) count of galaxy pairs separated by distance $r$, and $RR(r)$ is the expected pair count if galaxies were drawn from a uniform distribution.
$\xi(r)$ depends only on the HOD parameters that govern galaxy occupation; the IA parameters $\mucen$ and $\musat$ do not affect $\xi(r)$, as they influence only galaxy orientations.

We pre-compute all galaxy neighbor pairs $(i, j)$ with separations $|\mathbf{r}_{ij}| < r_{\rm max}$ using a KD-tree with periodic boundary conditions. The weighted pair count in each bin is computed as:
\begin{equation}
DD(r_k) = \sum_{(i,j) \in \mathcal{B}_k} w_i \, w_j,
\end{equation}
where $w_i$ and $w_j$ are the galaxy occupation weights and $\mathcal{B}_k$ denotes pairs with separations falling in bin $k$. For the expected pair count, we use the analytic expectation:
\begin{equation}
RR(r_k) = \left[ \left(\sum_i w_i\right)^2 - \sum_i w_i^2 \right] \frac{V_k}{V_{\rm box}},
\end{equation}
where $V_k = \frac{4\pi}{3}(r_{k+1}^3 - r_k^3)$ is the shell volume and $V_{\rm box}$ is the simulation volume. The term in brackets represents the effective number of distinct galaxy pairs for the weighted galaxy catalog.

While galaxy positions are fixed in \texttt{SubhaloPhaseSpace}, the weights $w_i$ are differentiable functions of the HOD parameters through $\langle \ncen \rangle$ and $\langle \nsat \rangle$. Gradients thus flow through the weight computation:
\begin{equation}
\frac{\partial \xi}{\partial \beta} = \sum_i \frac{\partial \xi}{\partial w_i} \frac{\partial w_i}{\partial \beta},
\end{equation}
where $\beta \in \{\log M_{\rm min}, \sigma_{\log M}, \log M_0, \log M_1, \alpha\}$.
Satellites placed via \texttt{NFWPhaseSpace} fallback are assigned weights proportional to $\langle \nsat \rangle$ of their host halo, preserving gradient flow.
This enables gradient-based inference of HOD parameters from galaxy clustering measurements.

We benchmark the accuracy of all differentiable correlation functions, including the IA correlations $\omega(r)$ and $\xi(r)$ on the fiducial HOD used in this work.
This is shown in Figure \ref{fig:diffcor}.
There is exact agreement between the \diffhodia and \halotoolsia estimators for $\omega(r)$, $\eta(r)$, and $\xi(r)$ in the case of unit galaxy weights.
In the case of HOD weights, which is necessary for full differentiability of $\xi(r)$ with respect to HOD parameters, the two correlations more noticeably differ, but still exhibit good agreement.

\section{Optimization Hyperparameters}
\label{app:hyperparams}

We provide details on the optimization hyperparameters used in the gradient-based experiments of Section~\ref{sec:experiments}.

\subsection{Moment-Matching Optimization}
\label{app:hyperparams_moment}

For the moment-matching objective described in Section~\ref{sec:momentmatching}, we optimize using the Adam optimizer \citep{kingma2017adammethodstochasticoptimization} with an initial learning rate of $5 \times 10^{-2}$ and an exponential decay schedule (decay rate 0.95 applied every 20 steps).
We run the optimization for 100 steps.
Gradients are clipped to a global norm of 10.0 to ensure numerical stability.
Parameters are initialized by sampling $\mucen$ and $\musat$ uniformly from $[-1, 1]$.
We repeat the optimization procedure 50 times with different random initializations.

\subsection{Correlation Function Optimization}
\label{app:hyperparams_omega}

For the correlation function objective described in Section~\ref{sec:correlationfunctionobjective}, we optimize $\mucen$ and $\musat$ using the Adam optimizer with an exponentially decaying learning rate schedule.
The initial learning rate is set to $5 \times 10^{-3}$, with a decay factor of 0.85 applied every 100 steps.
We run the optimization for 2000 steps.
Gradients are clipped to a global norm of 1.0 to prevent instabilities from large gradient magnitudes.
Parameters are constrained to $|\mu| < 1$ via clipping after each update.

The target correlation $\hat{\omega}(r)$ is computed by averaging over 20 orientation realizations of the fiducial \textsc{tng300} catalog, providing a low-noise reference.
We use 20 logarithmically-spaced radial bins spanning $0.1 < r < 16 \; h^{-1}\mathrm{Mpc}$.
The optimization catalog is generated with a different random seed than any of the target realizations, ensuring that we fit the statistical correlation structure rather than matching galaxy-by-galaxy.
As with the moment-matching case, we perform 50 independent optimization runs from random initializations uniformly sampled from $[-1, 1]$.

\end{document}